\documentclass[sn-aps]{sn-jnl}

\UseRawInputEncoding

\usepackage[utf8]{inputenc}

\jyear{2021}%

\theoremstyle{thmstyleone}%
\theoremstyle{thmstyletwo}%

\theoremstyle{thmstylethree}%

\begin{document}

\title[Thermal brachistochrone for harmonically confined Brownian particles]{Thermal brachistochrone for harmonically confined Brownian particles}


\author[1]{\fnm{Antonio} \sur{Patr\'on}}\email{apatron@us.es}

\author[1]{\fnm{Antonio} \sur{Prados}}\email{prados@us.es}

\author*[1]{\fnm{Carlos A.} \sur{Plata}}\email{cplata1@us.es}

\affil*[1]{\orgdiv{F\'isica Te\'orica}, \orgname{Universidad de Sevilla}, \orgaddress{Apartado de Correos 1065, E-41080 Sevilla, Spain}}


\abstract{The overdamped Brownian dynamics of a harmonic oscillator is a paradigmatic system in non-equilibrium statistical mechanics, which reliably models relevant stochastic systems such as colloidal particles submitted to optical confinement. In this work, optimal thermal protocols are tailored to minimise the connection time between equilibrium states of overdamped $d$-dimensional oscillators. Application of control theory reveals that these optimal protocols are of bang-bang type, that is, the temperature of the bath has to take alternatively the minimum and maximum values allowed. Minimum connection times increase with the considered dimension $d$. Remarkably, this is the case even for symmetric oscillators, for example, with spherical symmetry---in which the degeneracy of the elastic constant along the $d$ possible directions seems to imply a minimum connection time equal to that for the one-dimensional case. This surprising unavoidable price to pay when increasing dimension is thoroughly investigated and understood on a physical basis. Moreover, information theory tools such as the thermodynamic length and its divergence are analysed over the brachistochrone.}




\maketitle

\section{Introduction}\label{sec1}

Optimal shortcuts are a hot topic in physics. They were originally devised for quantum systems~\cite{chen_fast_2010,chen_shortcut_2010,guery-odelin_shortcuts_2019}, with the aim of driving the system of interest to the desired target state---beating its natural relaxation timescale, thus the name of shortcuts. Soon, these appealing ideas born in the quantum framework were extended and expanded to other contexts, such as classical mechanics and statistical mechanics~\cite{guery-odelin_nonequilibrium_2014,martinez_engineered_2016,patra_shortcuts_2017,li_shortcuts_2017,funo_shortcuts_2020}---see Ref.~\cite{guery-odelin_driving_2022} for a recent review. 

Shortcuts are based on the control of the time dependence of physical quantities that govern the dynamical evolution of the system at hand. Most often, one assumes that control parameters modulate the potential---for instance, the stiffness or center of an optical trap~\cite{schmiedl_optimal_2007,schmiedl_efficiency_2008,aurell_optimal_2011,aurell_boundary_2012}. Also, one can consider the whole potential as the control, which leads to quite general problems like the minimisation of irreversible work~\cite{aurell_optimal_2011,muratore-ginanneschi_extremals_2014,muratore-ginanneschi_application_2017,zhang_work_2020,zhang_optimization_2020} or the building of smooth connections that involve the manipulation of time's arrow~\cite{plata_taming_2021}. 

Only very recently have control problems in which the bath temperature is the control parameter been considered~\cite{martinez_adiabatic_2015,martinez_colloidal_2017,chupeau_thermal_2018,plata_finite-time_2020,plata_building_2020}. The interest in engineering the thermal environment stems from the design of novel experimental techniques that circumvent the difficulties for directly controlling the bath temperature. These techniques have opened new possibilities for the manipulation of Brownian objects and motivate the present work. For example, an effective increment of the temperature---up to thousands of kelvins---can be generated by submitting a colloidal particle to an additional white noise forcing. In particular, this can be achieved by applying a random electric field to charged particles~\cite{martinez_effective_2013}.

Optimal control problems that involve the engineering of the temperature have been scarcely addressed in the literature. A wide variety of physically relevant optimisation problems arise: minimisation of entropy production, irreversible work, thermodynamic length, connection time, et cetera---just as in the usual case of controlling the potential. In this paper, we focus on working out the thermal brachistochrone, that is, the temperature protocol that minimises the connection time between equilibrium states of the system, corresponding to different values of the temperature, $T_0$ and $T_f$. This problem
can be in principle tackled with the tools of optimal control theory, such as Pontryagin's maximum principle~\cite{pontryagin_mathematical_1987,liberzon_calculus_2012}. Typically, the temperature enters linearly in the evolution equations of the relevant physical quantities, which entails that the Pontryagin's Hamiltonian is linear in the control function---the temperature. When this is the case, the optimal thermal protocols that minimise the connection time between equilibrium states are of bang-bang type: they comprise several time intervals, where the control is constant and equals either its maximum or its minimum value. Protocols of this type have been previously reported in time-optimisation problems in different contexts, such as quantum systems~\cite{chen_fast_2010,ding_smooth_2020} or granular fluids~\cite{prados_optimizing_2021,ruiz-pino_optimal_2022}.

The main goal of this work is to analyse in depth the thermal brachistochrone for a $d$-dimensional harmonic oscillator. Not only does the harmonic oscillator have theoretical interest within the field of non-equilibrium statistical mechanics, but is also relevant from an experimental point of view: it accurately describes realistic systems such as confined colloidal particles~\cite{ciliberto_experiments_2017}. In fact, it is in the harmonic oscillator indeed that the first optimal control ideas were introduced within statistical physics~\cite{schmiedl_optimal_2007,aurell_optimal_2011,aurell_boundary_2012}, and it has also been employed to build irreversible heat engines~\cite{schmiedl_efficiency_2008,blickle_realization_2012,martinez_brownian_2016,martinez_colloidal_2017,plata_building_2020,nakamura_fast_2020}.

We consider that the temperature $T$ of the thermal bath can be externally controlled at will within a certain interval, $T_{\min}\leq T \leq T_{\max}$. This corresponds to a relevant experimental situation, as already stated above the temperature can only be increased from ``room'' temperature to thousands of kelvins~\cite{martinez_effective_2013}. Since our focus is put on thermal control, the elastic constants of the potential---there are $d$ of them---are kept constant. In the context of heat engines, protocols like those considered in this paper, in which the elastic constants are kept constant, correspond to isochoric branches.

In this work, we show that it is possible to explicitly work out the thermal brachistochrone for this paradigmatic case. We look into the dependence of the optimal connection on the limit values of the temperature, and also on the elastic constants that characterise the harmonic confinement.\footnote{In Ref.~\cite{prados_optimizing_2021}, the limiting case of a thermal bath with infinite heating power was briefly analysed.} A rich phenomenology emerges, including striking discontinuities of the minimum connection time when two or more of the elastic constants become equal---what we call the degenerate case. This means, for instance, that a perfect isotropic harmonic well and a slightly anisotropic one have very different minimum connection times: longer by a finite amount for the anisotropic well---however small the anisotropy is. It must be stressed that anisotropy in the harmonic confinement has been both experimentally observed and theoretically analysed  for laser optical tweezers~\cite{rohrbach_stiffness_2005,madadi_polarization-induced_2012,ruffner_universal_2014,yevick_photokinetic_2017,moradi_efficient_2019}.

The application of metrics and tools stemming from  information geometry to give insight into thermodynamic concepts is a fertile field of research, which dates back to the 80's of the past century~\cite{salamon_thermodynamic_1983,salamon_length_1985} and has exploded in the last years~\cite{crooks_measuring_2007,sivak_thermodynamic_2012,ito_stochastic_2018,nicholson_nonequilibrium_2018,ito_stochastic_2020,wadia_solution_2022}. Specifically, in relation to optimal control, information geometry allows for the derivation of classical speed limits~\cite{ito_stochastic_2018,shiraishi_speed_2018,nicholson_timeinformation_2020,ito_stochastic_2020}. These investigations prove that measures such as thermodynamic length and its divergence---sometimes also called thermodynamic cost~\cite{ito_stochastic_2018,ito_stochastic_2020}---are useful to improve our current understanding of non-equilibrium processes, including optimal shortcuts. Our work also discusses these novel concepts for the optimal thermal shortcuts derived.

The rest of the work is organised as follows. In section \ref{sec:model}, the dynamics of the model system, the $d$-dimensional harmonic oscillator, is introduced in detail within the framework of non-equilibrium statistical mechanics. Then, the problem of minimising the connection time between equilibrium states is posed and solved in sections \ref{sec:Tmax-fin} and \ref{sec:Tmax-inf} for two situations,  finite and infinite heating power, respectively. In section \ref{sec:inf-geo}, we resort to information geometry concepts to shed some light on the unavoidable extra cost for connecting equilibrium states of oscillators when the dimension $d$ is increased. A brief recap of the main results in this work, along with the conclusions that are extracted from them, is provided in section \ref{sec:concl}. Finally, some technical details are relegated to the appendices.

\section{The Model}\label{sec:model}

We consider an overdamped Brownian particle in $d$ dimensions in contact with a thermal bath at temperature $T(t)$ submitted to harmonic confinement in each direction.\footnote{The case $d>3$ corresponds to several, noninteracting, confined Brownian particles.} The origin is chosen to coincide with the center of the trap. Let $x_i$ be the projection of the position of the particle $\mathbf{x}$ onto the $i$-th dimension, while $k_i$ stands for the stiffness of the confining potential, $U(\mathbf{x})=\sum_{i=1}^{d}k_i x_i^2/2$, in the corresponding direction. For the sake of concreteness, we order the dimensions in such a way that $k_1 \leq k_2 \leq \cdots\leq k_d$. The temperature of the bath is externally controlled at will in such a way that it is possible to devise any temperature program---for an experimental implementation of a time-dependent bath temperature, see e.g. Ref.~\cite{martinez_effective_2013}. The stochastic dynamics can be cast either in the Langevin equation
\begin{equation}
\label{Langevin-equation}
\gamma \frac{d}{dt}\mathbf{x}(t) =- \nabla U(\mathbf{x}(t))+ \sqrt{2 \gamma k_B T(t)} \boldsymbol{\eta}(t),
\end{equation}
or the Fokker-Planck equation for the probability density function $P(\mathbf{x},t)$
\begin{equation}
\label{Fokker-Planck-equation}
\gamma \frac{\partial}{\partial t}  P(\mathbf{x},t) = \nabla \cdot [\nabla U (\mathbf{x}) P(\mathbf{x},t)] + k_B T(t) \nabla^2 P(\mathbf{x},t).
\end{equation}  
Above, $\gamma$ is the friction coefficient between the Brownian particle and the bath, $k_B$ stands for the Boltzmann constant, and $\boldsymbol{\eta}$ is a vector of white Gaussian noises, fully determined by the first two moments of its components
\begin{equation}
\langle \eta_i(t) \rangle = 0, \quad \langle \eta_i(t) \eta_j(t') \rangle = \delta (t-t') \delta_{ij}, \quad \forall(i,j).
\end{equation} 
These Gaussian noises are responsible for the diffusion term in the description of $P(\mathbf{x},t)$. Note that our assumption of uncoupled confinements in each dimension implies no loss of generality since, for general harmonic potentials, it is always possible to change variables to normal modes where our description applies---see Appendix~\ref{ap:normal-modes} for further details.

The linearity of the potential guarantees that Gaussian states remain Gaussian for the whole evolution of the system. Therefore, provided a Gaussian initial condition centered at the origin, which ensures $\langle x_i(t)\rangle=0$, it suffices to study the variance of the distribution in each dimension, $z_i(t) \equiv \langle x_i^2 (t)\rangle$, $i=1,\ldots,d$, to fully characterise the dynamics. Due to the uncoupling of the harmonic confinements, the evolution equations for the $z_i$ are also uncoupled and we have that 
\begin{equation}
\label{eq:Gaussian-shape}
P(\boldsymbol{x},t)=\prod_{i=1}^d \frac{e^{-\frac{x_i^2}{2z_i(t)}}}{\sqrt{2\pi z_i(t)}}  , \qquad \gamma \frac{d}{dt} z_i (t) = -2 k_i z_i(t) + 2 k_B T(t), \quad \forall i, \forall t. 
\end{equation} 
For simplifying the future discussion, it comes in handy to introduce dimensionless variables as
\begin{equation}
    t^* = \frac{k_1}{\gamma}t, \quad z_i^* = \frac{k_1}{k_B T_0}z_i, \quad T^* = \frac{T}{T_0}, \quad k_i^* = \frac{k_i}{k_1},
\end{equation}
with $T_0$ the initial temperature.
From now on, the asterisks are dropped to avoid cluttering notation. Hence, we have the evolution equations 
\begin{equation}
\label{eq:ev_z}
\frac{d}{dt} z_i (t) = -2 k_i z_i(t) + 2 T(t), 
\end{equation}      
with $k_1=1\leq k_2 \leq \cdots \leq k_d$. The system of equations \eqref{eq:ev_z} constitutes the fundamental law that governs the dynamics of the Brownian particle.

In this work, we are interested in devising a temperature program that allows the connection between two equilibrium states at different temperatures, $T(0)=T_0=1$ and $T(t_f)=T_f$, in the shortest amount of time. To the best of our knowledge, this optimisation problem---finding the thermal brachistochrone---is first posed and addressed in this work.   Since the control appears linearly in the dynamical equations, the optimal control will be a bang-bang protocol, analogously to the optimal protocols in the context of granular media~\cite{prados_optimizing_2021,ruiz-pino_optimal_2022} already mentioned in the introduction. Therefore, it is useful to solve these equations for constant $T$. 

The general solution of \eqref{eq:ev_z}, starting from the initial condition $z_{i,0}$ at $t_0$,  follows an exponential relaxation
\begin{equation}
\label{eq:ev-op}
\mathcal{E}_{k_i,T}^{(\Delta t)} \left( z_{i,0}\right) \equiv z_i(t)= \frac{T}{k_i} + \left( z_{i,0} - \frac{T}{k_i} \right) e^{-2k_i \Delta t}.
\end{equation} 
The evolution operator $\mathcal{E}_{k_i,T}^{(\Delta t)}$, with $\Delta t\equiv t-t_0$, generates the time evolution for a time interval $\Delta t$ of the $i$-th variance $z_i$ under constant temperature $T$.

\section{Optimal thermal protocols for finite heating power}
\label{sec:Tmax-fin}

Our goal is to obtain the protocol that connects two equilibrium states in the shortest time.  Specifically, in our dimensionless variables, the initial temperature is $T_0=1$, and thus final values $T_f$ higher (lower) than unity represent heating (cooling) processes. The solution to this optimisation problem depends on the constraints considered for the external control, that is, on the constraints on the temperature $T(t)$. A physical bound arises from below since temperature cannot be lower than zero. However, technical limitations may produce the emergence of tighter bounds in such a way that the control cannot exceed certain minimum and maximum values, that is, $0 \leq T_{\min} \leq T \leq T_{\max}$. In this section, we focus on cases with $T_{\min}=0$ and finite $T_{\max}$, whereas in Sec. \ref{sec:Tmax-inf} the latter condition will be relaxed: therein, we  assume an infinite heating power, that is, the limit as  $T_{\max}\to\infty$. Of course, one could consider a nonzero value for $T_{\min}$, and we put forward indeed a general approach, although thereafter we take $T_{\min}=0$ for the sake of simplicity in the presentation of the results.

The linearity of the equations guarantees the existence of the solution for the optimisation problem \cite{pontryagin_mathematical_1987}. 
The optimal protocol connects the initial and target states in the shortest (finite) time.  From a theoretical perspective, this is a qualitatively different from the direct step process where at $t=0^+$ the control is switched to the target value for the temperature, $T_f$, which is followed by an exponential relaxation of the variances with natural timescales given by $(2k_i)^{-1}$. From an applied perspective, the optimal protocol is especially appealing when the optimal time beats the aforementioned characteristic timescales.  

Pontryagin's maximum principle provides us with a perfect tool to address optimal problems submitted to constraints, such as the one we have posed~\cite{pontryagin_mathematical_1987,liberzon_calculus_2012}. In our case, the object to minimise is the total time employed in the process, which can be thought of as the simplest  functional  $t_f=\int_0^{t_f} dt$. This fact, together with the temperature appearing linearly in the evolution equations, makes Pontryagin's Hamiltonian linear in the control. Hence, optimal control has to be of the bang-bang type, meaning that $T(t)$ equals either $T_{\min}$ or $T_{\max}$---possibly with jumps in between---for all times $0 \leq t \leq t_f$. Specifically, the number of jumps between bounds is $d-1$, entailing $d$ stages in the interval $0<t<t_f$, and thus given by the dimensionality of the problem, as argued in Ref.~\cite{prados_optimizing_2021}. 
In this work, we explore in depth the physical implications of this general result in the relevant context of harmonically trapped Brownian  particles. 
In the following, we resort to our knowing the optimal control being of bang-bang type to obtain both the protocol itself and the optimal connection time. A detailed derivation of the optimal bang-bang protocol in our system for $d=2$, using explicitly Pontryagin's principle, can be found in Appendix~\ref{ap:pontryagin}.

\begin{figure} 
\begin{center}
\includegraphics[width=\textwidth]{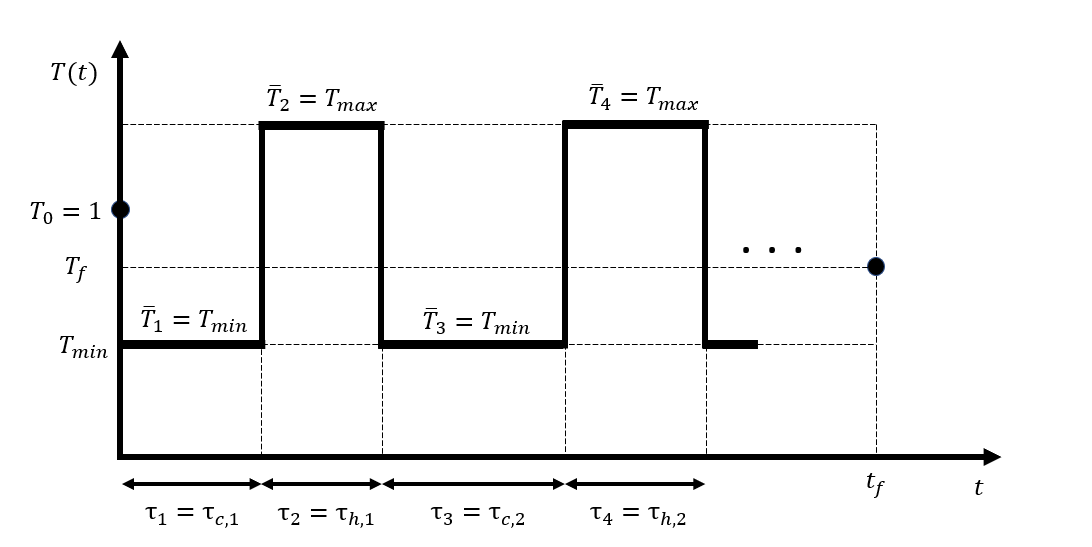}
\includegraphics[width=\textwidth]{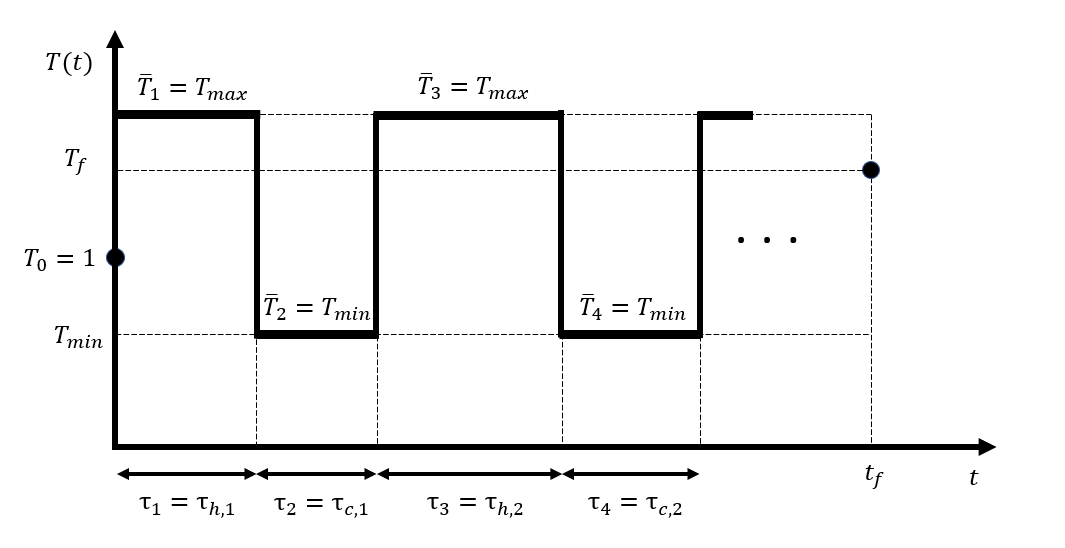}
\caption{\label{fig:sketch-Tmax-fin} Sketch of the optimal bang-bang control able to connect two equilibrium states of a $d$-dimensional harmonic oscillator. The optimal control comprises several bangs, that is, time windows with alternating maximum and minimum values of the bath temperature. The character of the first bang---the value of its bath temperature $\overline{T}_1$---is determined by the target temperature $T_f$ being either lower (cooling process, top panel) or higher (heating process, bottom panel) than the initial one $T_0=1$. The different parameters involved in the protocol, namely temperatures and time duration of the bangs, along with the notation used in Eq.~\eqref{eq:phi-fin} are displayed. The shortest connection time $t_f= \sum_{i=1}^d \tau_i$ stems from the solution of Eq.~\eqref{eq:comp}. }
\end{center}
\end{figure}

As discussed above, the shortest connection implies $d$ time windows, with $d-1$ consecutive jumps between the limiting values of the temperature. The value of the temperature along the first time window, either $T_{\max}$ or $T_{\min}$, determines the type of process performed: heating or cooling, respectively. For instance, let us consider a heating process, $T_f>1$: the optimal protocol involves $d$ time windows, alternating heating (at $T_{\max}$) and cooling (at $T_{\min}$) stages, starting with heating at $T_{\max}$. Note that, since the system is assumed to be at equilibrium both at the initial and final times, the boundary conditions $z_{i}(0)=1/k_i$ and $z_{i}(t_f)=T_f/k_i$ for the evolution hold. Let $\tau_{c,i}$ ($\tau_{h,i}$) be the duration of the $i$-th cooling (heating) stage. Therefore, the optimal bang-bang process is obtained by solving the system of equations
\begin{equation}
\label{eq:comp-h}
\underbrace{\left( \cdots \circ \mathcal{E}_{k_i,T_{\max}}^{\tau_{h,2}} \circ \mathcal{E}_{k_i,T_{\min}}^{\tau_{c,1}} \circ \mathcal{E}_{k_i,T_{\max}}^{\tau_{h,1}} \right)}_{{\scriptsize \textnormal{composition of }} d {\scriptsize \textnormal{ operators }}}
\left(\frac{1}{k_i}\right)=\frac{T_f}{k_i}, \quad i=1,\ldots,d
\end{equation} 
for heating processes, $T_f>1$, or
\begin{equation}
\label{eq:comp-c}
\underbrace{\left( \cdots \circ \mathcal{E}_{k_i,T_{\min}}^{\tau_{c,2}} \circ \mathcal{E}_{k_i,T_{\max}}^{\tau_{h,1}} \circ \mathcal{E}_{k_i,T_{\min}}^{\tau_{c,1}} \right)}_{{\scriptsize \textnormal{composition of }} d {\scriptsize \textnormal{ operators }}}
\left(\frac{1}{k_i}\right)=\frac{T_f}{k_i}, \quad i=1,\ldots,d
\end{equation} 
for cooling processes, $T_f<1$. Multiplying~\eqref{eq:comp-h} and \eqref{eq:comp-c} by $k_i$, we get
\begin{equation}
\label{eq:comp}
\varphi(k_i,T_{\min},T_{\max},\boldsymbol{\tau})=T_f, \qquad i=1,\ldots,d,
\end{equation}
where the function $\varphi$ can be easily built using the definition of the evolution operator \eqref{eq:ev-op} and $\boldsymbol{\tau}$ is a vector of dimension $d$ comprising all durations of the elemental stages in the bang-bang protocol. Note that, denoting by $\overline{T}_i$ (alternatively $T_{\max}$ or $T_{\min}$) and $\tau_i$ the temperature and the duration of the $i$-th elementary stage, respectively, of the bang-bang process, the function $\varphi(k,T_{\min},T_{\max},\boldsymbol{\tau})$ can be generally expressed as 
\begin{equation}
\label{eq:phi-fin}
\varphi(k,T_{\min},T_{\max},\boldsymbol{\tau}) =  \sum_{n=1}^{d+1} \left( \overline{T}_{n-1} - \overline{T}_n\right) \exp \left( -2 k \sum_{m=n}^{d}\tau_m \right),
\end{equation}
for arbitrary $d$. We have defined $\overline{T}_0\equiv 1$ and $\overline{T}_{d+1}\equiv 0$ to give a compact formulation. For an illustration of the optimal control protocol, and the notation employed for our general formulation in Eq.~\eqref{eq:phi-fin}, see Fig.~\ref{fig:sketch-Tmax-fin}.

If all the elastic constants $k_i$ are different, the mathematical problem is completely closed: Eq.~\eqref{eq:comp} is a system of $d$ equations for $d$ unknowns, the components of the vector $\boldsymbol{\tau}$ of time spans. The optimal time for the connection is given by the sum of all components of the solution for $\boldsymbol{\tau}$. 
Despite the formulation of the mathematical problem being simple, the system of  equations~\eqref{eq:comp} is difficult to solve for arbitrary dimension. Therefore, we discuss in the following physically meaningful situations within relevant limits. 

Now, let us consider the case when some $k_i$, say $r$ of them, are exactly equal---which we refer to as the degenerate case. This is relevant from a physical point of view, since it arises naturally when the harmonic confinement possesses some symmetry, for example cylindrical or spherical in the three-dimensional case. In such a situation, the variances corresponding to dimensions with the same elastic constant fulfill the same mathematical relationship. In such a situation, the problem can be solved by considering that the dimension has been reduced to $d^*=d+1-r<d$, i.e. $d^*$ is the number of different values of the elastic constant $k_i$. The optimal protocol  would thus involve a smaller number $d^*$ of elementary stages or bangs. 

Now, let us address the case in which the $r$ $k_i$'s are arbitrarily close, but not exactly equal. Therein, we need the  $d$ bangs to achieve the optimal connection and the situation is quite subtle, as shown below. For the sake of concreteness, we study the almost fully degenerate case $r=d$, defined as the limit where all confinements are almost identical:  $k_i \to k_1=1$, $\forall i>1$. This describes a harmonic trap with almost spherical symmetry in $d$ dimensions. This is an experimentally relevant situation, since the elastic constants along orthogonal directions would not be perfectly equal in a real experiment~\cite{rohrbach_stiffness_2005,madadi_polarization-induced_2012,ruffner_universal_2014,yevick_photokinetic_2017,moradi_efficient_2019}. Our use of the word ``almost'' stresses the fact that the $k_i$'s are not exactly equal.

For the almost fully degenerate case, the system of equations Eq.~\eqref{eq:comp} is no longer closed if  we set all the $k_i$'s equal to unity: there are $d$ unknowns but only one equation.
Nevertheless, it is possible to resort to a perturbative approach that helps us find the missing $d-1$ equations required to close the problem.\footnote{If only some  $k_i$'s were almost equal, there would be $d$ unknowns and $d^*<d$ equations. In that case, a similar perturbative approach would provide us with the missing equations.} By setting $k_i=1+\epsilon_i$, with $\epsilon_i\ll 1$ for $i>1$, one gets
\begin{subequations}
\begin{eqnarray}
\label{eq:comp-deg1}
\varphi(1,T_{\min}, T_{\max},\boldsymbol{\tau}) &=& T_f,
 \\
\label{eq:comp-deg2}\frac{\partial^n}{\partial k^n} \varphi(k,T_{\min}, T_{\max},\boldsymbol{\tau})\biggr\rvert_{k=1} &=& 0, \quad n=1,\ldots,d-1.
\end{eqnarray}
\end{subequations}

The solution of \eqref{eq:comp-deg1} and \eqref{eq:comp-deg2} for the vector of time spans $\boldsymbol{\tau}$ provides us with the optimal bang-bang protocol in the almost fully degenerate case. Below, we show that this problem does not converge to the solution of the fully degenerate case, that is, to the one-dimensional solution. This is a remarkable property of the system under study: increasing the dimension comes at an an unavoidable price, in which the shortest connection time presents a jump when going from $d$ to $d+1$. This happens even when the elastic constants are almost equal in all directions, and the confinement is arbitrarily close to be spherically symmetric. 

Below, we look into the solution for optimal connections for $d=1$ and the almost fully degenerate case for $d=2$ and $d=3$. Not only is this done to be concrete, but also because it is an experimentally relevant situation for one colloidal particle trapped in an almost isotropic harmonic trap. In each case, we mainly compute the function $\varphi(k,0,T_{\max},\boldsymbol{\tau})$ for both cooling, $T_f<1$, and heating, $T_f>1$, and discuss the results that stem from the found solution. We keep a finite $T_{\max}$ as a parameter but, as previously introduced, we choose $T_{\min}=0$ for the sake of simplicity.

\subsection{One-dimensional case}
\label{subsec:1-d-Tmax-fin}
Considering first a cooling process, $T_f<1$, it is simple to get
\begin{equation}
\varphi(k,0, T_{\max},\boldsymbol{\tau}) =e^{-2 k t_f}
\end{equation}
for $d=1$, where $\boldsymbol{\tau}$ is just the duration of the cooling process, $t_f=\tau_{c,1}$. Directly applying Eq.~\eqref{eq:comp} and solving for $t_f$, we obtain the shortest cooling time  
\begin{equation}
\label{eq:tmin_1d}
t_f=-\frac{1}{2} \ln T_f
\end{equation}
associated to the final temperature $T_f$---we recall that  $k_1=1$. Since there is no heating stage for the fastest cooling protocol in the one-dimensional case, Eq.~\eqref{eq:tmin_1d} is independent of $T_{\max}$.\footnote{Therefore, it coincides with the result for the limit $T_{\max}\to\infty$ in Ref.~\cite{prados_optimizing_2021}.} The protocol leading to the shortest time to cool down the particle is reasonable from a physical point of view: put the system in contact with a thermal bath at zero temperature and wait until the target state is reached. Consistently, the connection time diverges when $T_f \to 0$, and monotonically decreases up to zero for $T_f \to 1 $; a qualitative behaviour that is expected for all $d$.

For heating processes, $T_f>1$, we obtain
\begin{equation}
\varphi(k,0, T_{\max},\boldsymbol{\tau})=  \left(1- T_{\max} \right) e^{- 2 k t_f} + T_{\max},
\end{equation}
for $d=1$, where the only heating stage of duration $t_f$ represents the whole process, $t_f=\tau_{h,1}$. Solving Eq.~\eqref{eq:comp} for $t_f$ in this case yields
\begin{equation}
t_f= \frac{1}{2} \ln \left(\frac{T_{\max}-1}{T_{\max}-T_f} \right).
\end{equation}
Once more, the qualitative behaviour can be intuitively justified. The shortest time is identically zero just for $T_f=1$. It increases with $T_f$ until it diverges  for $T_f \to T_{\max}$.  Furthermore, in the limit $T_{\max} \to \infty$ the optimal time vanishes for all $T_f>1$. This describes an instantaneous heating process, consequence of the infinite capacity to heat up. This limit will be further investigated in Sec. \ref{sec:Tmax-inf}. 

\begin{figure}
\flushright\includegraphics[width=0.875\textwidth]{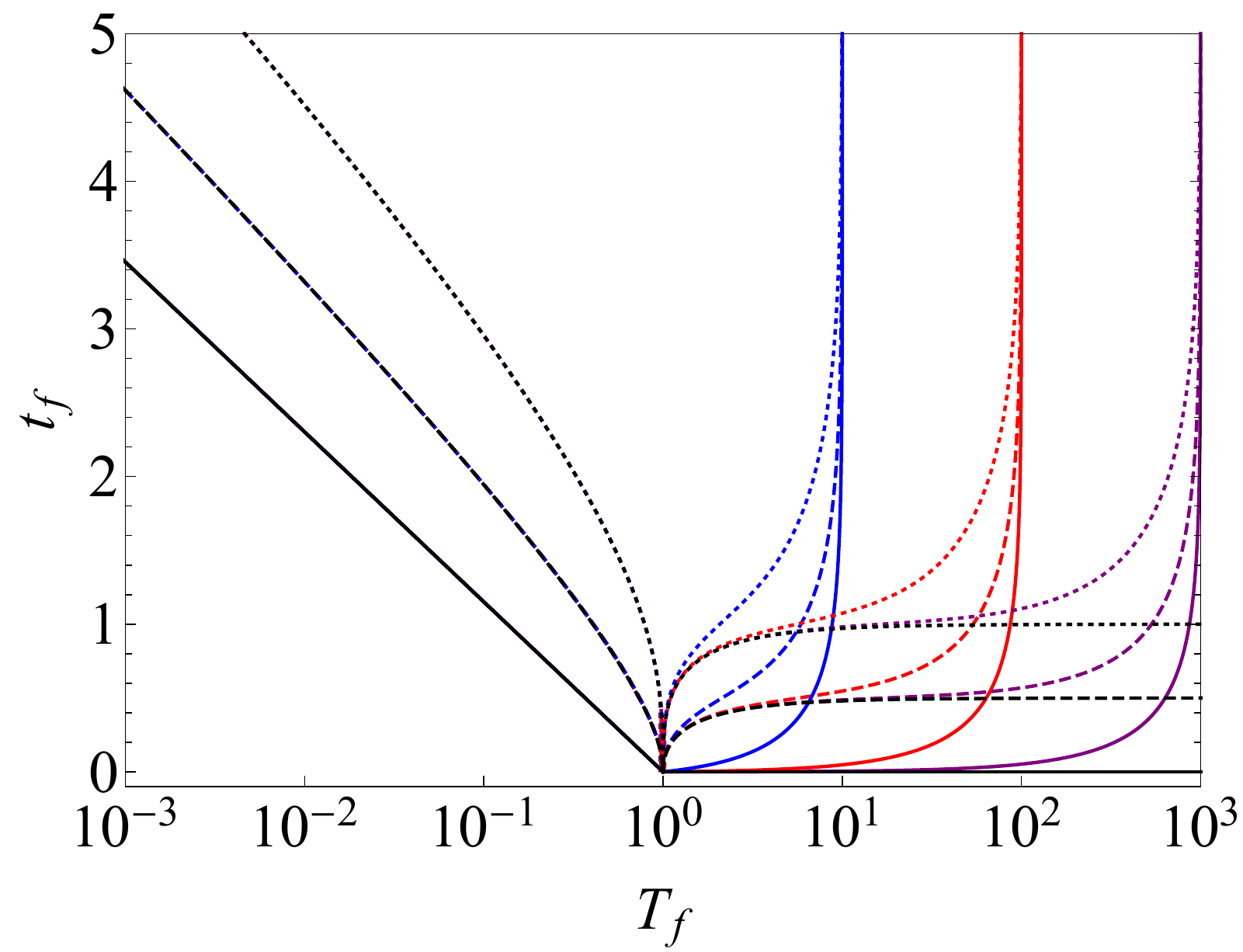}
\caption{\label{fig:tf_vs_Tf_difTmax}Shortest connection time as a function of the final temperature for different heating powers $T_{\max}$ and dimensions $d$. Different colours stand for the different values of $T_{\max}$ considered, namely $T_{\max}=10$ (blue), $T_{\max}=100$ (red), $T_{\max}=1000$ (purple), and $T_{\max} \to \infty$ (black). Different dimensions are displayed with different patterns: $d=1$ (solid), $d=2$ (dashed) and $d=3$ (dotted). At the scale of the plot, the results in the cooling region ($T_f<1$) are hardly distinguishable from the limit behaviour  $T_{\max} \to \infty$ given by the black curve. For heating processes, the universal behaviour corresponding to $T_{\max} \to  \infty$ is approached in the regime $T_f/T_{\max}\ll 1$.}
\end{figure}

The results obtained for $d=1$ are shown in Fig.~\ref{fig:tf_vs_Tf_difTmax} with solid lines. Therein, also plotted are (i) the results obtained for the two-dimensional (dashed lines) and three-dimensional (dotted lines) cases, which are obtained in Secs. \ref{subsec:2-d-Tmax-fin} and \ref{subsec:3-d-Tmax-fin}, and (ii) the results for the limit as $T_{\max} \to \infty$, which are derived in Sec. \ref{sec:Tmax-inf}. A detailed comparison between these results is done later, in the corresponding sections of the paper.

\subsection{Two-dimensional case}
\label{subsec:2-d-Tmax-fin}

For $d=2$, two bangs or elementary processes are needed. Therefore, the maximum (minimum) value of the temperature plays a role even for cooling (heating). As shown in the following, the fact of tuning two variances to their corresponding target states implies an increment of the cost of the optimal process, in terms of the minimum connection time to achieve the connection---even in the almost degenerate case. Hence, the optimal connection time in a system with effective $d=2$, for arbitrary $k_2$,  will be larger or equal than the corresponding one for $d=1$. On the one hand, for finite $k_2$, equality only holds for the trivial value $T_f=1$ which represents no change in the system and it is of course instantaneous. On the other hand, the convergence to the results corresponding to $d=1$ can be also recovered when the limit $k_2 \to \infty$ is considered. Therein, the trap along the second spatial direction is completely rigid and thus always in equilibrium with vanishing variance. Such a limit will be further analysed in Sec. \ref{sec:Tmax-inf} for the case of infinite heating power, $T_{\max} \to \infty$. 

Now we show that the convergence to the one-dimensional case discussed above for $k_2\to\infty$ is not present for other physically relevant limits, where one would naively expect it to hold: the almost degenerate case $k_2 \to k_1=1$.
Analytic solutions of $\boldsymbol{\tau}$ for the system of equations \eqref{eq:comp-deg1}--\eqref{eq:comp-deg2} are not possible for $d>1$. For the sake of clarity, we provide in the following the function $\varphi$ concerned in both situations, cooling and heating.\footnote{The derivatives involved in Eq.~\eqref{eq:comp-deg2} are easy to compute given the simple structure of $\varphi$. Nevertheless, since the final system is not especially illuminating and  cannot be worked out much further analytically, we do not write it down explicitly.} In the cooling case, $T_f<1$, we get
\begin{equation}
\varphi(k,0, T_{\max},\boldsymbol{\tau})=  e^{- 2 k (\tau_1+\tau_2)} -T_{\max}e^{- 2 k\tau_2}+T_{\max},
\end{equation}
whereas for heating, the function is
\begin{equation}
\label{two-dim-heating}
\varphi(k,0, T_{\max},\boldsymbol{\tau})= \left(1-T_{\max} \right) e^{- 2 k (\tau_1+\tau_2)} +T_{\max}e^{- 2 k\tau_2}.
\end{equation}
The pair $(\tau_1,\tau_2)$ represents $(\tau_{c,1},\tau_{h,1})$ for cooling and  $(\tau_{h,1},\tau_{c,1})$ for heating.

In Fig.~\ref{fig:tf_vs_Tf_difTmax}, the final optimal time $t_f=\tau_1+\tau_2$ is evaluated for the numerical solution of the system of equations \eqref{eq:comp-deg1}--\eqref{eq:comp-deg2} and displayed with dashed lines. The vertical asymptotes at $T_{\min}=0$ (cooling branch of the $t_f$ vs.~$T_f$ curve) and $T_{\max}$ (heating branch) are preserved. As briefly described in Sec. \ref{subsec:1-d-Tmax-fin}, going from $d=1$ to $d=2$ entails a finite increment in the minimum connection time---even in the almost degenerate situation. This unexpected asymmetry is one of the main results of our work and entails an unavoidable price when controlling higher-dimensional systems, even if they are almost isotropic.

\begin{figure}
\includegraphics[width=\textwidth]{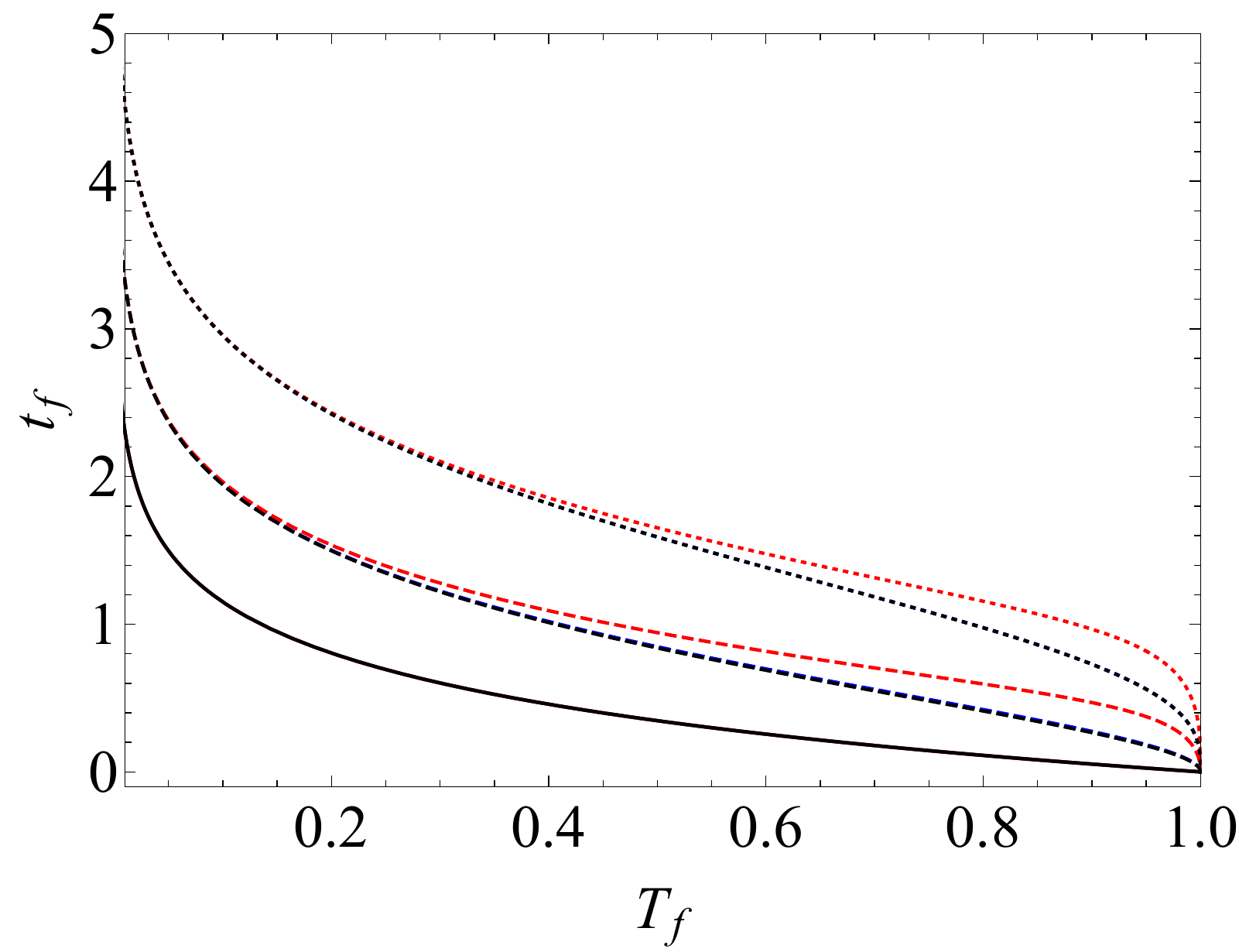}
\caption{\label{fig:tf_vs_Tf_difTmax2} Zoom of the cooling branch of the  $t_f$ versus $T_f$ curve. Different heating powers $T_{\max}$ and dimensions $d$ are considered with the following code: $T_{\max}=10$ (blue), $T_{\max}=1.1$ (red), $T_{\max} \to \infty$ (black), and the same code as in Fig.~\ref{fig:tf_vs_Tf_difTmax} for $d$. Blue and black curves are always indistinguishable. The scale and the particular case $T_{\max}=1.1$ shown here make it possible to discern the effect of $T_{\max}$ on $t_f$ over the cooling branch, which was imperceptible in Fig.~\ref{fig:tf_vs_Tf_difTmax}.}
\end{figure}

In the heating branch, $t_f$ monotonically increases with $T_f$. Remarkably, in the intermediate regime $1<T_f<T_{\max}$, we find a convergence to a universal behaviour. The asymptote $t_f=0.5$ arising in the limit as $T_{\max}\to \infty$ is studied in detail in Sec. \ref{sec:Tmax-inf}. In the cooling branch, the shown results evidence that $T_{\max}$ does not significantly influence the minimum connection time.  This is a reasonable property, given that---for the whole range of cooling---the ratio $T_{\max}/T_f$ is relatively big and the heating stage is thus expected to be short. In Fig.~\ref{fig:tf_vs_Tf_difTmax2}, we present results for the cooling branch with a quite low value of $T_{\max}=1.1$, in order to show that $T_{\max}$ indeed affects the shortest connection time. 

In the above, we have focused on the limit $k_2\to k_1=1$, but the non-degenerate case can also be considered. In such a situation, the minimum connection time monotonically decreases with $k_2$, tending to that of the one-dimensional case in the limit as $k_2 \to \infty$. The dependence on $k_2$ is investigated in more detail within the limit $T_{\max}\to \infty $ in Sec. \ref{sec:Tmax-inf}. 

\subsection{Three-dimensional case}
\label{subsec:3-d-Tmax-fin}

For a three-dimensional system, three elementary stages or bangs are required. The discussion is similar to the one carried out for the two-dimensional case, but with an additional jump. Herein, we study the effect of adding a new dimension, which involves an extra cost in terms of the optimal connection time.  The limit $k_3 \to \infty$ recovers the results for the two-dimensional case. However, this is not the case for $k_3 \to k_2$. For the sake of concreteness, we study in detail the almost fully degenerate case, where the limits $k_i\to k_1=1$ for $i=2,3$ are introduced. Also analogously to the two-dimensional case, we numerically solve the system of equations~\eqref{eq:comp-deg1}--\eqref{eq:comp-deg2}.

In the following, the functions $\varphi$ are provided. For cooling, $T_f<1$, one gets
\begin{equation}
\varphi(k,0, T_{\max},\boldsymbol{\tau})=  e^{- 2 k (\tau_1+\tau_2+\tau_3)} -T_{\max}e^{- 2 k(\tau_2+\tau_3)}+T_{\max}e^{-2k \tau_3}.
\end{equation}
while for heating, it is
\begin{eqnarray}
\varphi(k,0, T_{\max},\boldsymbol{\tau})&=& \left(1-T_{\max} \right) e^{- 2 k (\tau_1+\tau_2+\tau_3)} +T_{\max}e^{- 2 k(\tau_2+\tau_3)} \nonumber \\
&& -T_{\max}e^{-2k\tau_3}+T_{\max}.
\end{eqnarray}
The triplet $(\tau_1,\tau_2,\tau_3)$ stands for $(\tau_{c,1},\tau_{h,1},\tau_{c,2})$ in the cooling case and for $(\tau_{h,1},\tau_{c,1},\tau_{h,2})$ in the heating one.

The results of $t_f=\tau_1+\tau_2+\tau_3$ for the solution of  
the system of equations \eqref{eq:comp-deg1}--\eqref{eq:comp-deg2} are shown in Fig.~\ref{fig:tf_vs_Tf_difTmax} with dotted lines. A relevant part of the qualitative behaviour for $d=1$ and $d=2$ is preserved: presence of the same vertical asymptotes, same monotonicity for cooling and heating.  Also, going from $d=2$ to $d=3$ entails a finite increment of the minimum connection time, as was the case when going from $d=1$ to $d=2$. The emergence of a universal minimum connection time in the limit as $T_{\max} \to \infty$ will be analysed in Sec. \ref{sec:Tmax-inf}. As in the two-dimensional case, for observing the influence of the upper bound $T_{\max}$ on the minimum time for the cooling branch it is necessary to consider low values of $T_{\max}$, see Fig.~\ref{fig:tf_vs_Tf_difTmax2}.

\section{Optimal thermal protocols for infinite heating power}
\label{sec:Tmax-inf}

In this section, we study the same time optimisation problem for the $d$-dimensional harmonically trapped Brownian particle, but assuming that there is no upper bound for the temperature of the heat bath, that is, $T_{\max}\to \infty$. A priori, one could expect this limit to be singular. Nonetheless, the infinite heating power entails a vanishing time for the heating bangs, $\tau_{h,i} \to 0$. In turn, this entails  that $T_{\max} \tau_{h,i}$ must tend to a certain finite constant, which becomes a new unknown that, in what follows, plays the role that $\tau_{h,i}$ had earlier.

The limit $T_{\max}\to\infty$ could be directly applied to equations \eqref{eq:comp-deg1}--\eqref{eq:comp-deg2} but, for the sake of clarity, we choose to introduce it in the evolution operator from the very beginning. To this end, the finite constants
\begin{equation}
\alpha_{h,i} \equiv 2 T_{\max} \tau_{h,i}
\end{equation} 
are defined. In this limit, the evolution operator in Eq.~\eqref{eq:ev-op} is 
\begin{equation}
\widetilde{\mathcal{E}}_{\alpha_h} \left(z_{i,0}\right)= \lim_{T_{\max}\to \infty \atop 2T_{\max}\tau_{h} \to \alpha_h} \mathcal{E}_{k_i,T_{\max}}^{\tau_{h}} \left(z_{i,0}\right)= z_{i,0}+ \alpha_h,
\end{equation}
which no longer depends on the elastic constants $k_i$---as explicitly stated in our notation for $\widetilde{\mathcal{E}}$.
The equivalent relations to those in Eqs.~\eqref{eq:comp-h} and \eqref{eq:comp-c} are thus
\begin{equation}
\label{eq:comp-h-inf}
\underbrace{\left( \cdots \circ \widetilde{\mathcal{E}}_{\alpha_{h,2}} \circ \mathcal{E}_{k_i,T_{\min}}^{\tau_{c,1}} \circ \widetilde{\mathcal{E}}_{\alpha_{h,1}} \right)}_{{\scriptsize \textnormal{composition of }} d {\scriptsize \textnormal{ operators }}}
\left(\frac{1}{k_i}\right)=\frac{T_f}{k_i}, \quad i=1,\ldots,d
\end{equation} 
for heating processes, $T_f>1$, and
\begin{equation}
\label{eq:comp-c-inf}
\underbrace{\left( \cdots \circ \mathcal{E}_{k_i,T_{\min}}^{\tau_{c,2}} \circ \widetilde{\mathcal{E}}_{\alpha_{h,1}} \circ \mathcal{E}_{k_i,T_{\min}}^{\tau_{c,1}} \right)}_{{\scriptsize \textnormal{composition of }} d {\scriptsize \textnormal{ operators }}}
\left(\frac{1}{k_i}\right)=\frac{T_f}{k_i}, \quad i=1,\ldots,d
\end{equation} 
for cooling processes, $T_f<1$. Similarly to Eq.~\eqref{eq:comp}, it is possible to give a compact form for Eqs.~\eqref{eq:comp-h-inf} and \eqref{eq:comp-c-inf},
\begin{equation}
\label{eq:comp-inf}
\widetilde{\varphi}(k_i,T_{\min},\boldsymbol{\tau}_c,\boldsymbol{\alpha}_h)=T_f, \quad i=1,\ldots,d,
\end{equation}
where the vectors $\boldsymbol{\tau}_c$ and $\boldsymbol{\alpha}_h$ contain  the durations of the cooling stages and the intensity of the heating ones, respectively. The sum of the dimension of both vectors is equal to $d$ while the absolute value of their difference is zero (unity) for even (odd) $d$.

\begin{figure} 
\begin{center}
\flushright
\includegraphics[width=0.875\textwidth]{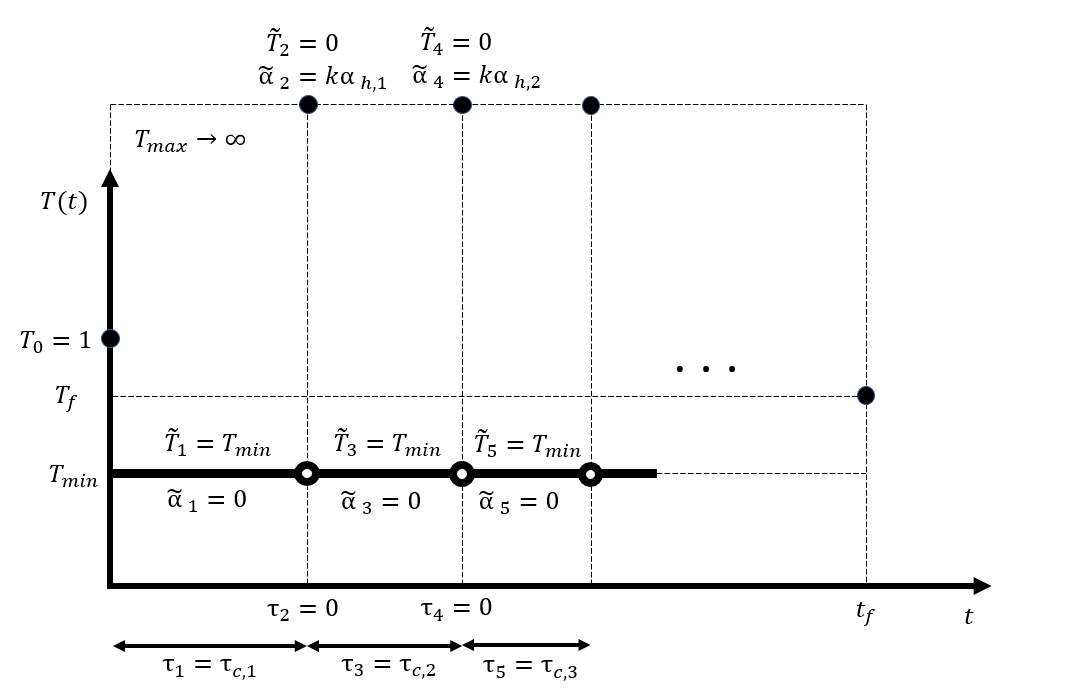}
\includegraphics[width=0.875\textwidth]{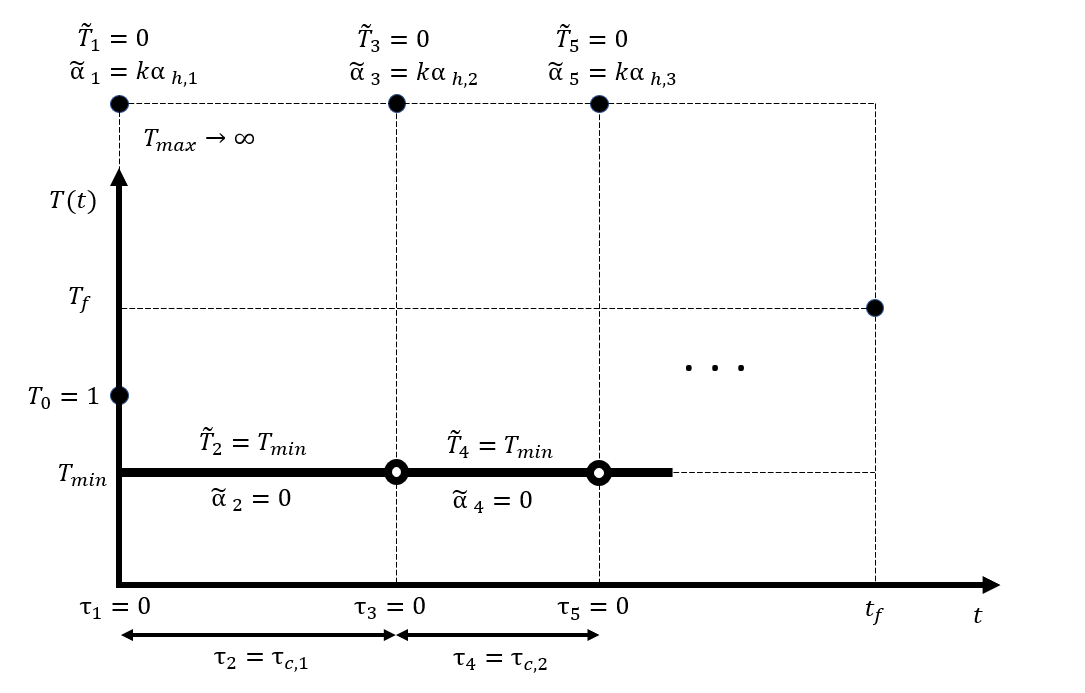}
\caption{\label{fig:sketch-Tmax-inf} Sketch of the optimal bang-bang control connecting the initial and final states, for infinite heating power $T_{\max}\to\infty$. The alternating elementary bangs are either cooling stages of finite duration or instantaneous heating stages. The top (bottom) panel corresponds to cooling (heating), that is, $T_f <1$ ($T_f>1$).
The different parameters involved in the protocol, namely temperatures, heating jumps and time spans of the elementary bangs, along with the notation used in Eq.~\eqref{eq:phi-inf}, are displayed. The shortest connection time $t_f= \sum_{i=1}^d \tau_i$ stems from the solution of Eq.~\eqref{eq:comp-inf}. Note that $t_f=\tau_{c,1}+\tau_{c,2}+\cdots$, i.e. only the cooling bangs contribute to the minimum connection time. }
\end{center}
\end{figure}

It is possible to give a general expression for $\widetilde{\varphi}(k,T_{\min},\boldsymbol{\tau}_c,\boldsymbol{\alpha}_h)$, specifically
\begin{equation}
\label{eq:phi-inf}
\widetilde{\varphi}(k,T_{\min},\boldsymbol{\tau}_c,\boldsymbol{\alpha}_h) =  \sum_{n=1}^{d+1} \left( \widetilde{T}_{n-1} + \widetilde{\alpha}_n - \widetilde{T}_n\right) \exp \left( -2 k \sum_{m=n}^{d}\tau_m \right), \quad \forall d,
\end{equation}
where:
\begin{enumerate}
\item $\widetilde{T}_i$ takes values either zero or $T_{\min}$ depending on whether the $i$-th stage is of heating or of cooling, respectively.
\item $\widetilde{T}_0=1$ and $\widetilde{T}_{d+1}=\widetilde{\alpha}_{d+1}=0$.
\item $\widetilde{\alpha}_i$ takes values either sequentially from $\boldsymbol{\alpha}_{h}$ multiplied by $k$ or zero depending on whether the $i$-th stage is of heating or of cooling, respectively.
\item $\tau_i$ takes values either zero or sequentially from $\boldsymbol{\tau}_{c}$ depending on whether the $i$-th stage is of heating or of cooling, respectively.
\end{enumerate}  
Thus, for the non-degenerate case, the problem of searching the optimal protocol minimising the connection time is reduced to solving Eq.~\eqref{eq:comp-inf}, with $\widetilde{\varphi}$ given by Eq.~\eqref{eq:phi-inf}. A sketch for the optimal control protocol, and the notation employed in our general formulation in Eq.~\eqref{eq:phi-inf} is presented in Fig.~\ref{fig:sketch-Tmax-inf}

For the almost fully degenerate case, one can proceed along similar lines to those for finite $T_{\max}$---developed in Sec. \ref{sec:Tmax-fin}. A perturbative approach, again considering that $k_i=1+\epsilon_i$, $\epsilon_i\ll 1$, makes it possible to close the mathematical problem for the $d$ unknowns in the $\boldsymbol{\alpha}$ and $\boldsymbol{\tau}$ vectors. Specifically, we obtain
\begin{subequations}
\begin{eqnarray}
\label{eq:comp-deg1-inf}
\widetilde{\varphi}(1,T_{\min},\boldsymbol{\tau}_c,\boldsymbol{\alpha}_h) &=& T_f, \\
\label{eq:comp-deg2-inf}\frac{\partial^n}{\partial k^n} \widetilde{\varphi}(k,T_{\min},\boldsymbol{\tau}_c,\boldsymbol{\alpha}_h)\biggr\rvert_{k=1} &=& 0, \quad n=1,\ldots,d-1.
\end{eqnarray}
\end{subequations}
In the following, the specific forms of the function $\widetilde{\varphi}$ are worked out for $1\leq d\leq 3$. 
 
\subsection{One-dimensional case}

For a cooling process, there is no difference with the case of finite $T_{\max}$. Specifically, it is straightforward to obtain 
\begin{equation}
\widetilde{\varphi}(k,0,\boldsymbol{\tau}_c,\boldsymbol{\alpha}_h) = e^{-2k t_f},
\end{equation}
 where $\tau_1=\tau_{c,1}=t_f$ is the minimum connection time---the duration of the single cooling bang in the protocol. Hence, the minimum time Eq.~\eqref{eq:tmin_1d} is reobtained.

For a heating process, the limit $T_{\max}\to \infty$ indeed induces a change, since
\begin{equation}
\widetilde{\varphi}(k,0,\boldsymbol{\tau}_c,\boldsymbol{\alpha}_h) = 1+k\alpha
\end{equation}
In this occasion, Eq.~\eqref{eq:comp-inf} just gives the correct value of $\alpha_1=\alpha_{h,1}$ to reach the target state.  There is only one heating stage, which is instantaneous $t_f=\tau_{h,1}=\alpha/2T_{\max}=0$. Reasonably, the minimum time to perform a heating process vanishes since one has infinite resources to heat up the system. As we are about to demonstrate, this is no longer the case for $d>1$.  

The results for the minimum connection time in the limit as $T_{\max} \to \infty $ are shown in Figs.~\ref{fig:tf_vs_Tf_difTmax} and \ref{fig:tf_vs_Tf_difTmax2} with solid black lines.

\subsection{Two-dimensional case}

Here, the optimal protocol comprises two bangs or elementary stages, as in the case of finite $T_{\max}$. The main difference stems from  the heating bangs being instantaneous. The role of the duration of the heating $\tau_{h,1}$ is now played by its intensity $\alpha_{h,1}=2T_{\max}\tau_{h,1}$. Using the general formula in Eq.~\eqref{eq:phi-inf}, one gets
\begin{equation}
\widetilde{\varphi}(k,0,\boldsymbol{\tau}_c,\boldsymbol{\alpha}_h) = e^{-2kt_f}+k\alpha
\end{equation}
for a cooling process, $T_f<1$, while 
\begin{equation}
\widetilde{\varphi}(k,0,\boldsymbol{\tau}_c,\boldsymbol{\alpha}_h) =(1+k \alpha) e^{-2 k t_f}
\end{equation}
for a heating process, $T_f>1$. The pairs  $(\widetilde{\alpha}_1,\widetilde{\alpha}_2)$ and $(\tau_1,\tau_2)$  stand for $(0,k \alpha_{h,1} = k \alpha)$ and $(\tau_{1,c}=t_f,0)$ in the cooling process and for $(k \alpha_{h,1}=k \alpha,0)$  $(0,\tau_{c,1}=t_f)$ in the heating process, respectively,. 

\begin{figure}
\includegraphics[width=\textwidth]{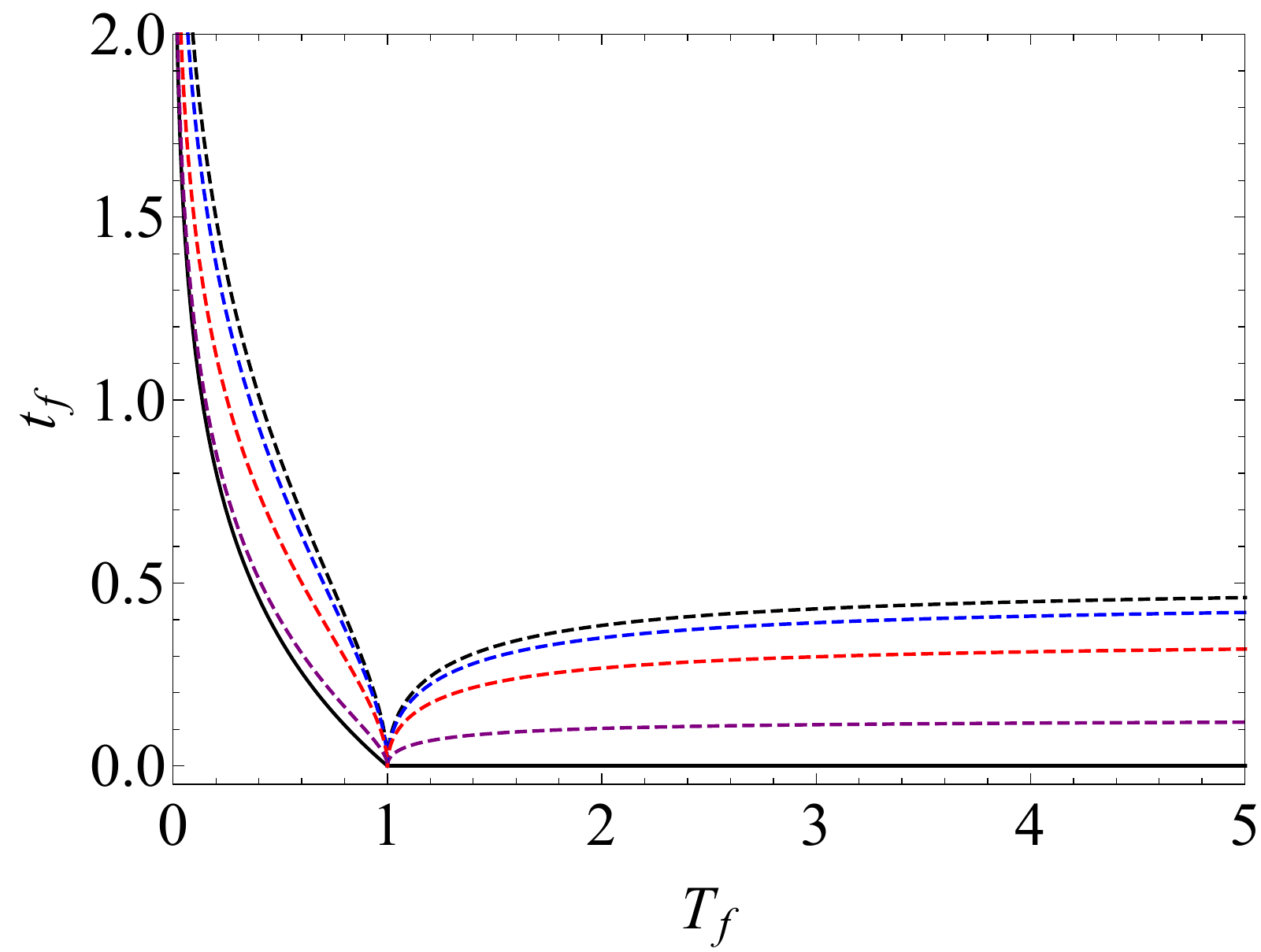}
\caption{\label{fig:tf_vs_Tf_difk} Shortest connection time as a function of the final temperature, for different elastic constants  $k_2$ and dimensions $d$. Different colors stand for the different values of $k_2$ considered, namely $k_2=1.1$ (blue), $k_2=5$ (red), $k_2=10$ (purple) and $k_2 \to 1$ (black). Different dimensions follow the same code as in previous figures, that is, results for $d=1$ ($d=2$) correspond to solid (dashed) lines.}
\end{figure}

Solving Eqs.~\eqref{eq:comp-inf} for $k_2$ is possible up to reach the inverse relation between the optimal time and the final temperature, that is, $T_f$ may be expressed as an analytic function of $t_f$.  These formulas are not particularly illuminating for our purposes.\footnote{They can be found in Ref.~\cite{prados_optimizing_2021}.} Instead, we discuss the dependence of the resulting $t_f$ with the second elastic constant $k_2$. On the one hand, as already discussed in Sec. \ref{sec:Tmax-fin}, the limit of infinite confinement, $k_2 \to \infty$, entails that the second spatial direction becomes irrelevant---its spatial variance identically vanishes for all times---and the resulting $t_f$ converges to that for $d=1$. On the other hand, the almost degenerate limit, $k_2 \to 1$, is subtler. There is no possibility of dimensional reduction and solving the system of two unknowns given by Eqs.~\eqref{eq:comp-deg1-inf} and \eqref{eq:comp-deg2-inf} is mandatory. One has that
\begin{equation}\label{eq:tfinv_Tmaxinf_2d_c}
T_f= e^{-2t_f} (1+2t_f),
\end{equation}
fog cooling, and 
\begin{equation}
\label{eq:tfinv_Tmaxinf_2d_h}
T_f=\frac{e^{-2t_f}}{1-2t_f},
\end{equation}
for heating. 
The curves corresponding to Eqs.~\eqref{eq:tfinv_Tmaxinf_2d_c} and \eqref{eq:tfinv_Tmaxinf_2d_h} are displayed in Figs.~\ref{fig:tf_vs_Tf_difTmax} and \ref{fig:tf_vs_Tf_difTmax2} with dashed black lines. Note that Eq.~\eqref{eq:tfinv_Tmaxinf_2d_h} predicts the emergence of the horizontal asymptote for  $t_f=1/2$ in  Fig.~\ref{fig:tf_vs_Tf_difTmax}. Moreover, we illustrate in Fig.~\ref{fig:tf_vs_Tf_difk} the dependence of $t_f$ on the second elastic constant $k_2$. Therein, the behaviour found above for the physically relevant limits $k_2 \to \infty$ and $k_2 \to 1$ is manifest. Moreover, it is clearly observed that, at fixed target temperature $T_f$, the minimum connection time  monotonically decreases with $k_2$. 

As highlighted several times before, there appears a time cost associated with increasing the dimension of the system. This feature is especially remarkable in the limit as $T_{\max} \to  \infty$: the minimum time needed to achieve the connection for the two-dimensional case is finite, even when having an infinite heating power---at variance with the one-dimensional case, for which $t_f=0$. In this respect, it is convenient to recall that the fastest cooling rate is reached when the temperature of the thermal bath takes its minimum value, which is  physically bounded by zero. This forbids instantaneous cooling processes, that is, the cooling power is always limited---which explains the observed asymmetry between cooling and heating processes.  In Appendix~\ref{ap:Tmin-inf}, this bound is disregarded by assuming an unphysical scenario, where $T_{\min} \to -\infty$. Therein, we show that it is  finite cooling power that finite minimum connection times stem from.

\subsection{Three-dimensional case}
Now, three bangs or elementary stages are involved in the optimal protocol. There are three unknowns, two $\alpha_h$'s and one $\tau_c$ for heating processes, and vice versa for cooling processes. Using the general formula in Eq.~\eqref{eq:phi-inf}, we get
\begin{equation}
\widetilde{\varphi}(k,0,\boldsymbol{\tau}_c,\boldsymbol{\alpha}_h) = e^{-2k (\tau_{c,1}+\tau_{c,2})} + k\alpha e^{-2k \tau_{c,2}},
\end{equation}
for a cooling process, whereas
\begin{equation}
\widetilde{\varphi}(k,0,\boldsymbol{\tau}_c,\boldsymbol{\alpha}_h)= (1 +k \alpha_{h,1})e^{-2k t_f} + k\alpha_{h,2}.
\end{equation}
for a heating process. The triplets  $(\widetilde{\alpha}_1,\widetilde{\alpha}_2,\widetilde{\alpha}_3)$ and $(\tau_1,\tau_2,\tau_3)$  stand, respectively, for $(0,k \alpha_{h,1} = k \alpha,0)$ and $(\tau_{1,c},0,\tau_{2,c})$ in the cooling process and for $(k \alpha_{h,1},0,k \alpha_{h,2})$ and  $(0,\tau_{c,1}=t_f,0)$ in the heating one.

Again, an implicit expression for $t_f$ in terms of $T_f$ can be obtained, for arbitrary $k_2$ and $k_3$ for both heating and cooling---which, once more, are not especially illuminating.\footnote{They can also be found in Ref.~\cite{prados_optimizing_2021}.}  Below, we focus on the almost fully degenerate case, which is interesting from a physical point of view: an imperfect spherical symmetry.  Starting from  Eqs.~\eqref{eq:comp-deg1-inf} and \eqref{eq:comp-deg2-inf} for $d=3$, it is possible to derive that
\begin{equation}
\label{eq:tfinv_Tmaxinf_3d_c}
T_f= e^{-2t_f} \left[ 1+2t_f^2 + 2t_f \sqrt{ (1+t_f^2)} \right],
\end{equation}
fog cooling, and 
\begin{equation}
\label{eq:tfinv_Tmaxinf_3d_h}
T_f=\frac{e^{-2t_f}(1+t_f)}{1-t_f},
\end{equation}
for heating. 
The curves corresponding to Eqs.~\eqref{eq:tfinv_Tmaxinf_3d_c} and \eqref{eq:tfinv_Tmaxinf_3d_h} have already been represented in Figs.~\ref{fig:tf_vs_Tf_difTmax} and \ref{fig:tf_vs_Tf_difTmax2} with dotted black lines. The horizontal asymptote at $t_f=1$ predicted by \eqref{eq:tfinv_Tmaxinf_3d_h} is evident in Fig.~\ref{fig:tf_vs_Tf_difTmax}.

\subsection{Higher dimensions}
The approach introduced above can be carried out to address an almost fully degenerate case for arbitrary dimension $d$. This is not an unphysical problem: for example, it corresponds to $d$ colloidal particles, each one trapped in its own one-dimensional harmonic potential---with all the harmonic wells being almost equal. 

For heating processes, there always emerges an asymptotic value $t_f^{asy} \equiv \lim_{T_f \to \infty} t_f$ for the minimum connection time.  This value is an upper bound for the minimum connection time for a finite value of the target temperature $T_f>1$ of a heated system, $t_f(T_f)\leq t_f^{asy}$. The bound $t_f^{asy}$ monotonically increases with the dimension, presenting finite jumps when going from $d$ to $d+1$---as already discussed for the particular cases $d=1\to 2$ and $d=2\to 3$. In Table~\ref{tab:asym}, the values for the optimal time for these asymptotes are displayed. These values have been analytically computed solving Eqs.~\eqref{eq:comp-deg1-inf} and \eqref{eq:comp-deg2-inf}, except for $d=6$, which has been numerically obtained. Note that, for  $1\leq d\leq 3$, $t_f^{asy}$ follows the  simple formula $t_f^{asy}=(d-1)/2$: this simple expression is broken for $d=4$.
\begin{table}
\caption{\label{tab:asym} Values for the asymptotic value of the minimum heating time $t_f^{asy}$ in the almost fully degenerate case for a  $d$-dimensional oscillator. Finding the reported values involve solving a polynomial equation of degree $d-1$: therefore, for $d>5$, the solution has been obtained numerically. }

\begin{center}
\begin{tabular}{cc}
\hline
\multicolumn{1}{l}{Dimension $d$} & \multicolumn{1}{c}{
$t_f^{asy}$} \\ \hline
1                     & 0                                            \\
2                     & $\frac{1}{2}$                                \\
3                     & 1                                            \\
4                     & $\frac{2+\sqrt{2}}{2}$                       \\
5                     & $\frac{3+\sqrt{3}}{2}$                       \\
6                     & 3.14497                                      \\ \hline
\end{tabular}
\end{center}
\end{table}

\section{Information geometry}
\label{sec:inf-geo}

We have shown the emergence of an unavoidable price when adding dimensions to the almost degenerate oscillator under consideration, in terms of the time that has to be paid in order to make the shortest connection.  In this section, we study this expense from the point of view of information geometry, an enticing field with a great potential to assist non-equilibrium statistical mechanics~\cite{crooks_measuring_2007,sivak_thermodynamic_2012,amari_information_2016,ito_stochastic_2018,ito_stochastic_2020,nicholson_timeinformation_2020}. Specifically, the thermodynamic length $\mathcal{L}$ and its divergence $\mathcal{C}$---sometimes also called the thermodynamic cost~\cite{ito_stochastic_2018,ito_stochastic_2020}---are thoroughly analysed for the optimal bang-bang protocols along this work.     

A key quantity in the information geometry framework is the Fisher information
\begin{equation}
\label{eq:Fisher-def}
    I(t) = \int d\boldsymbol{x} \  \left( \partial_t \ln P(\boldsymbol{x},t) \right)^2 P(\boldsymbol{x},t).
\end{equation}
From it, one can write down expressions for the so-called statistical or thermodynamic length
\begin{equation}
\label{eq:length-def}
    \mathcal{L}(t) = \int_0^{t}ds \ \sqrt{I(s)},
\end{equation}
which measures the length of the path swept by the system in probability space, and also the thermodynamic cost or Fisher divergence
\begin{equation}
\label{eq:cost-def}
    \mathcal{C}(t) =\frac{1}{2} \int_0^t ds \ I(s).
\end{equation}
In the calculation of the Fisher information, the Gaussian nature of the distribution plays a remarkably simplifying role. In particular, introducing Eq.~\eqref{eq:Gaussian-shape} into Eq.~\eqref{eq:Fisher-def} leads to
\begin{equation}
\label{eq:Fisher-gaus}
    I(t) = \frac{1}{2}\sum_{i=1}^d \left[\frac{\dot{z}_i(t)}{z_i(t)}\right]^2,
\end{equation}
where we have made use of $\int d\boldsymbol{x} \  x^2_i x^2_j P(\boldsymbol{x},t) = (1+2\delta_{i,j}) z_i(t)z_j(t)$.

In the fully degenerate case, all terms in the sum in Eq.~\eqref{eq:Fisher-gaus} are identical and thus the Fisher information simplifies to
\begin{equation}
\label{eq:Fisher-deg}
    I(t) = \frac{d}{2} \left[\frac{\dot{z}(t)}{z(t)}\right]^2.
\end{equation}
This expression is also valid for the almost fully degenerate case, for which $k_i=1+ \epsilon_i$, with $\epsilon_i\ll 1$, since the difference with Eq.~\eqref{eq:Fisher-deg} vanishes in the limit $\epsilon_i\to 0$, $\forall i$. A subtlety should be remarked, though: when evaluating the Fisher information over the optimal protocol, the path swept by the system in probability space---codified in the time evolution of $z(t)$---is different for the fully degenerate and the almost fully degenerate cases. For the former, the optimal path is identical to that for the one-dimensional case, which comprises only one bang. For the latter, the optimal path comprises $d$ bangs, with the upper and lower bounds of the temperature alternating over it.

\subsection{Thermodynamic length}\label{sec:thermo-length}

Henceforth, we restrict ourselves to the almost fully degenerate case---consistently, our starting point is   Eq.~\eqref{eq:Fisher-deg}. First, let us note that  
\begin{equation}\label{eq:length-inequal}
    \mathcal{L}(t) =  \sqrt{\frac{d}{2}}\int_0^t ds \ \biggr\rvert\frac{\dot{z}(s)}{z(s)}\biggr\rvert \geq \sqrt{\frac{d}{2}}\biggr\rvert \ln \frac{z(t)}{z(0)}\biggr\rvert \;\Longrightarrow\; \mathcal{L}(t_f) \geq  \sqrt{\frac{d}{2}}\rvert \ln T_f \rvert,
\end{equation}
which bounds $\mathcal{L}(t_f)/\sqrt{d}$ by its optimal value for $d=1$. Now, we derive $\mathcal{L}(t_f)$ for the optimal bang-bang protocol in arbitrary $d$ dimensions. For the sake of simplicity, and consistently with our approach in Secs. \ref{sec:Tmax-fin} and \ref{sec:Tmax-inf}, we set $T_{\min}=0$. Taking into account the exponential relaxation in the cooling bangs, described by the evolution operator $\mathcal{E}_{1,0}^{\tau_c}$, as given by Eq.~\eqref{eq:ev-op}, one gets
\begin{equation}
\label{eq:length-Tmax}
 \mathcal{L}(t_f) = \sqrt{\frac{d}{2}} \left[ \ln T_f + 4 \tau_{c}\right],
\end{equation}
where $\tau_c$ is the total time employed in cooling stages. Equation~\eqref{eq:length-Tmax} holds for arbitrary values of $T_{\max}$, being the assumptions $T_{\min}=0$ and almost fully degeneration the only hypotheses necessary for deriving it. Notably, the limit $T_{\max} \to \infty$ further simplifies Eq.~\eqref{eq:length-Tmax}, since the heating bangs are instantaneous and thus $\tau_c=t_f$:
\begin{equation}
\label{eq:length-inf}
 \mathcal{L}(t_f) = \sqrt{\frac{d}{2}} \left[ \ln T_f + 4 t_f \right].
\end{equation}
Substituting Eq.~\eqref{eq:length-inf} into the inequality in Eq.~\eqref{eq:length-inequal} yields the following bounds for the connection time:
\begin{equation}
\label{bounds-1d}
    t_f\geq - \frac{1}{2}\ln T_f , \; T_f<1; \qquad  t_f \geq 0 , \; T_f>1.
\end{equation}
These bounds are precisely the values over the optimal connection for $d=1$.

Apart from the expected multiplicative factor $\sqrt{d}$ in Eq.~\eqref{eq:length-inf}, the thermodynamic length has an additional increment when going from $d$ to $d+1$ stemming from its dependence with $t_f$. This feature can be understood as a fingerprint of the aforementioned unavoidable cost of increasing the spatial dimension of the system, even in the almost fully degenerate case.

\subsection{Thermodynamic cost}

When computing the thermodynamic cost, the linearity in the integrand allows us to integrate each dimension separately, making it unnecessary the assumption of fully degenerate systems to derive an analytical expression for $\mathcal C$. By introducing the general formula~\eqref{eq:Fisher-gaus} into Eq.~\eqref{eq:cost-def}, we obtain 
\begin{equation}
\mathcal{C} (t_f) =  \sum_{i=1}^d \left[ - k_i \ln T_f + \int_0^{t_f} ds \,\dot{z}_i(s)\frac{T(s)}{z_i^2(s)} \right].
\end{equation}
We have changed to $z_i$ as variable of integration, employed Eq.~\eqref{eq:ev_z}, and taken into account the initial and final values variances, $z_i(0)=1/k_i$ and $z_i(0)=T_f/k_i$. The temperature $T(s)$ alternatively takes the extreme values $T_{\min}$ (cooling bangs) and $T_{\max}$ (heating bangs). For the case of our concern, the cost simplifies to
\begin{equation}
\label{cost}
\mathcal{C} (t_f) = - \sum_{i=1}^d \left[  k_i \ln T_f + T_{\max} \sum_j \Delta_{h,j} \left(\frac{1}{z_i}\right) \right],
\end{equation}
where $\Delta_{h,j} (1/z_i) $ refers to the change of the inverse of the $i$-th variance over the $j$-th heating bang---only the heating bang contributes to the sum when $T_{\min}=0$.  On the one hand, the first term, $-\sum_{i=1}^d k_i \ln T_f$ increases linearly with $d$ in the almost fully degenerate case. On the other hand, the second term is expected to increase faster than linearly with $d$, 
due to the double sum over $i$ and $j$.

\subsection{Speed limit}\label{sec:speed-limits}
Speed limits are bounds on the rate of evolution of dynamical systems. Although originally derived for quantum-mechanical systems---see Ref.~\cite{deffner_quantum_2017} for a recent review, lately they have been derived for classical systems with stochastic dynamics~\cite{ito_stochastic_2016,ito_stochastic_2018,shiraishi_speed_2018,nicholson_nonequilibrium_2018,plata_finite-time_2020,shiraishi_speed_2020,ito_stochastic_2020,nicholson_timeinformation_2020}. They involve trade-off relationships between the evolution speed, or the connection time between initial and final states, and information geometry quantities, which account for the cost required for driving the system 
or the length of the followed path---among others. For isothermal processes, these bounds can be related to physical quantities such as  irreversible work or  entropy production~\cite{sivak_thermodynamic_2012,nakazato_geometrical_2021}. For processes with varying temperature, like those considered in this paper, this connection is not straightforward and remains an open question, to the best of our knowledge.

Taking into account the definitions of  thermodynamic length and cost in Eqs.~\eqref{eq:length-def} and \eqref{eq:cost-def}, direct application of the Cauchy-Schwarz inequality entails that the connection time between two states must verify the inequality~\cite{ito_stochastic_2018,ito_stochastic_2020}
\begin{equation}\label{eq:speed-limit}
    t_f\geq t^{geo}\equiv \frac{\mathcal{L}^2}{2\mathcal{C}},
\end{equation}
where $t^{geo}$ is evaluated over the specific path swept by the system in the connection. The quantity $t^{geo}$ is thus an information geometry lower bound for the connection time. The inequality \eqref{eq:speed-limit} is saturated only over the geodesic, that is, $t_f=t^{geo}$ only for the path connecting the initial and target points that minimises $\mathcal{L}$. We remark that only both the heating and cooling procedures with $T_{\max} \rightarrow \infty$ and $T_{\min} = 0$ in the one dimensional case correspond to such geodesics, being the geodesic and the brachistochrone different in general.

\begin{figure}
\flushright\includegraphics[width=0.875\textwidth]{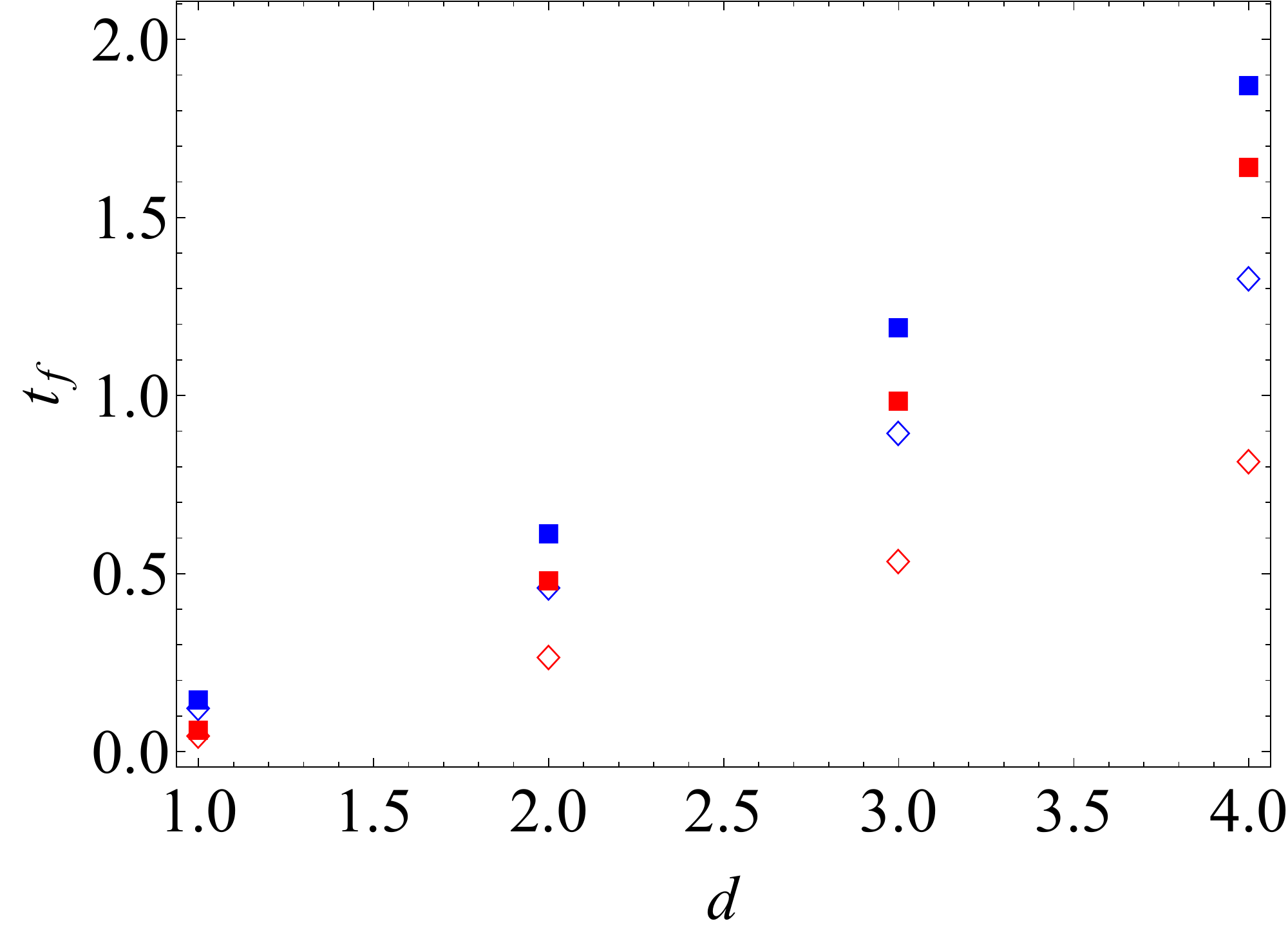}
\caption{Dimension dependence of both the optimal connection time $t_f$ (filled squares) and the geometric lower bound  $t^{geo}=\mathcal{L}^2/(2C)$ in Eq.~\eqref{eq:speed-limit} (empty diamonds) for a general, almost fully degenerate, harmonic oscillator. Two sets of data are shown for different values of $T_{\max}$: $T_{\max} = 5$ (blue) and $T_{\max} = 10$ (red). Additional parameters employed are $T_f = 2$ and $T_{\min}=0$.} \label{speed-limit-graph}
\end{figure}
As discussed before, both $\mathcal{L}/\sqrt{d}$ and $\mathcal{C}/d$ are expected to be increasing functions of $d$, due to the contribution thereto of the terms that involve details of the optimal connection---specifically,  the terms involving $t_f$ in Eq.\eqref{eq:length-inf} and $\sum_j\Delta_h (1/z_i) $ in Eq.\eqref{cost}, respectively. Thus, the quantity $t^{geo} = (\mathcal{L}/\sqrt{d})^2/(2\mathcal{C}/d)$,  is expected to have a non-trivial dependence on the dimension in the almost fully degenerate case we are considering. Therefore, it is worth investigating the dependence on $d$ of the geometric bound $t^{geo}$.  Figure \ref{speed-limit-graph} shows that $t^{geo}$ increases with $d$, which hints once more at the unavoidable cost of increasing the spatial dimension of the system. We must also point out that $t^{geo}$ decreases with $T_{\max}$, vanishing in the limit as $T_{\max} \rightarrow \infty$, while the optimal connection times $t_f$ tend to the finite values derived in Sec.~\ref{sec:Tmax-inf}.  For $T_{\max} \rightarrow \infty$, $\mathcal L$ remains finite and $\mathcal C$ diverges in such a limit, which implies that $t^{geo} \rightarrow 0$ for all $d$. Interestingly, this limit behaviour is similar to that found for a uniformly heated granular system in Ref.~\cite{prados_optimizing_2021}.

\section{Discussion}	
\label{sec:concl}

What is the fastest connection, that is, the brachistochrone, between two states of a mesoscopic physical system? This is a relevant question from a theoretical perspective, bringing to the fore concepts from the theory of stochastic processes~\cite{van_kampen_stochastic_1992}, stochastic thermodynamics~\cite{sekimoto_stochastic_2010,peliti_stochastic_2021}, and optimal control theory~\cite{pontryagin_mathematical_1987,liberzon_calculus_2012}. Also, it is a relevant question from an applied perspective: for example, minimum time protocols have been shown to be the building blocks of the adiabatic---in the sense of zero average heat---branches of a maximum power irreversible Carnot-like engine~\cite{plata_building_2020}. 

Our work gives an analytical solution to the above question for a paradigmatic model system, which is also significant for actual experiments---since it describes an optically trapped colloidal particle. This is a $d$-dimensional harmonic oscillator, the elastic constants of which are kept constant but the temperature of the bath in which it is immersed can be externally controlled.  Specifically, the system is initially at equilibrium at temperature $T_0$ and we want to drive it to a final equilibrium state with temperature $T_f$. It is important to stress that this is a relevant physical situation, which corresponds to isochoric (zero work) protocols. 

A first key result of our work is showing that the thermal brachistochrone is a protocol of the bang-bang type, which comprises alternating heating and cooling time windows with the maximum ($T_{\max}$) and minimum ($T_{\min}$) values available for the bath temperature. For the sake of simplicity, we have set $T_{\min}=0$ for the lower bound of the temperature---from a physical standpoint, this means that $T_{\min}\ll T_f$. It is worth stressing that our approach holds for both limited heating power, finite $T_{\max}$, that is, $T_f/T_{\max}=O(1)$, and infinite heating power $T_{\max} \to \infty$, that is, $T_{\max}\gg T_f$.

The bang-bang protocols derived here are significant because they allow for a fast---in fact, the fastest---and precise way to drive the system from its initial equilibrium state at the temperature $T_0$ to its final equilibrium state at the desired target temperature $T_f$. This is in contrast with the direct quench from $T_0$ to $T_f$, which requires an infinite amount of time to exactly reach the final equilibrium state. Moreover, we would like to highlight that the results for the brachistochrone obtained  here are exact; no approximation has been made, despite the problem being highly non-trivial. Also, the arguments leading to the emergence of the bang-bang protocols as those minimising the connection time are quite general, which hints at the possibility of extending the results presented here to more general systems---for example with non-harmonic confinement.

A second key result of our approach is the finite increment of the minimum connection time when moving from dimension $d$ to $d+1$, even when all the elastic constants are (almost) equal---what we have called the almost fully degenerate case. For a particle confined in a perfectly isotropic $d$-dimensional harmonic well---what we call the fully degenerate case, the minimum connection time equals that of the one-dimensional case. The evolution equations for the variances in all directions are identical and the $d$ degrees of freedom are effectively reduced to only one. However, in a real experiment, there appear differences in the elastic constants along the different directions and isotropy (or spherical symmetry) is thus not perfect~\cite{rohrbach_stiffness_2005,madadi_polarization-induced_2012,ruffner_universal_2014,yevick_photokinetic_2017,moradi_efficient_2019}. Intuitively, one expects these small differences to have a small, infinitesimal, impact in the minimum connection time. But our work shows that there appears a finite increment of the minimum connecting time, however small the anisotropy---i.e. the deviation from perfect spherical symmetry---is.

This intriguing and inescapable price for higher-dimensional slightly anisotropic, or almost fully degenerate, systems is unexpected and constitutes an outstanding result of our work. The finite increment of the connection time when going from dimension $d$ to $d+1$ stems from the optimal bang-bang protocol comprising as many stages as the number of different elastic constants, independently of the magnitude of the difference among them. To shed further light to this regard, we have resorted to information geometry concepts, such as the thermodynamic length and its divergence---also called thermodynamic cost~\cite{ito_stochastic_2020}. Not only do thermodynamic length and cost share the remarkable feature of additional expenses when considering systems with increasing dimensionality, but also the geometric time bound associated with them increases with dimension.

Our work opens several perspectives for future research, some of which we highlight in the following. First, it would be interesting to extend the ideas developed here to the underdamped regime. The Gaussian behaviour persists in the underdamped regime, which should allow for obtaining a simple dynamical system involving not only the variances of the position $\langle x^2 \rangle$ but also the rest of second moments, namely $\langle v^2 \rangle$ and $\langle x v \rangle$. Still, the possible existence of oscillatory modes---depending on the relative values of the natural frequencies and the damping constant---makes the problem non-trivial from the point of view of control theory. Second, the current technical development of optical trapping and the experimental techniques that make it possible to control the temperature of the bath in an effective manner allow for the actual implementation of the optimal protocols derived here in the laboratory. The instantaneous switchings of the temperature can be engineered by making the bath temperature vary over a time scale much shorter than that characterising the dynamical behaviour of the Brownian particle. We recall that the bath temperature can be effectively varied by applying a random electric field to a charged colloid~\cite{martinez_colloidal_2017}. Third, our results are remarkably relevant for devising optimal heat engines made of Brownian objects. Since the emergence of stochastic thermodynamics~\cite{sekimoto_stochastic_2010,peliti_stochastic_2021}, the goal of building functional Brownian heat engines has been a persistent aspiration that has been addressed from both theoretical and experimental perspectives~\cite{blickle_realization_2012,martinez_colloidal_2017,plata_building_2020,nakamura_fast_2020,zhang_optimization_2020,tu_abstract_2021}. For a harmonic oscillator, infinitesimal work is given by $\sum_i \left<x_i^2\right>dk_i/2$, where $\left<x_i^2\right>$ is the spatial variance and $k_i$ is the elastic constant in the $i$-th direction. Therefore, protocols with constant stiffnesses, as those corresponding to thermal shortcuts, have an identically vanishing work throughout the considered path and, in this sense, are analogous to isochoric  processes in ``traditional'' heat engines. Many classical thermodynamic cycles, such as Stirling's \cite{blickle_realization_2012,muratore-ginanneschi_efficient_2015,krishnamurthy_micrometre-sized_2016} and Otto's \cite{deng_boosting_2013,abah_shortcut--adiabaticity_2019}, comprise isochoric branches, which endows the results derived in this paper  with extra significance for future applications. Finally, it is worth investigating from a physical standpoint the consequences that the time-optimal paths designed here have on other relevant quantities of interest within the framework of stochastic thermodynamics, such as entropy production~\cite{muratore-ginanneschi_extremals_2014,landi_irreversible_2021} or irreversible work~\cite{aurell_optimal_2011,zhang_work_2020,zhang_optimization_2020}.

\bmhead{Acknowledgments}
We acknowledge financial support from Grant PGC2018-093998-B-I00 funded by MCIN/AEI/10.13039/501100011033/ and by ERDF ``A way of making Europe.'' C.A.P.  acknowledges financial support from Junta de Andaluc\'{\i}a and European Social Fund through the program PAIDI-DOCTOR. A.~Patr\'on acknowledges support from the FPU programme through Grant FPU2019-4110. Discussions with Ra\'ul A. Rica are also acknowledged.

\begin{appendices}

\section{General harmonic potential and normal modes}\label{ap:normal-modes}

In this appendix, our minimisation problem for a general harmonic potential is recast. We consider the same overdamped Brownian particle in $d$-dimensions in contact with a thermal bath at temperature $T(t)$. However, in this case, the particle is trapped inside a general harmonic potential of the form
$U(\mathbf{q}) = \sum_{i,j=1}^d \kappa_{ij}q_i q_j/2$, with $q_i$ being the $i$-th spatial coordinate in a certain basis, and $\kappa_{ij}$ being the elements of a symmetric, positive-definite matrix $\mathbf{K}$, which accounts for the stiffness of the trap. The corresponding stochastic descriptions are given by
\begin{equation}
\label{Langevin-equation-general}
    \gamma \frac{d}{dt}\mathbf{q}(t) = -\nabla U(\mathbf{q}(t)) + \sqrt{2\gamma k_B T(t)} \boldsymbol{\hat{\eta}}(t),
\end{equation}
and
\begin{equation}
\label{Fokker-Planck-equation-general}
    \gamma \frac{\partial}{\partial t}P(\mathbf{q},t) = \nabla \cdot [\nabla U (\mathbf{q}) P(\mathbf{q},t)] + k_B T(t) \nabla^2 P(\mathbf{q},t),
\end{equation}
with $\gamma$, $k_B$, $\boldsymbol{\hat{\eta}}$ and $P(\mathbf{q},t)$ having the same definitions as those at the beginning of Sec. \ref{sec:model}. We introduce now the normal modes $\mathbf{x}$,
\begin{equation}
\label{eq:change-qx}
    q_i = \sum_{j=1}^d C_{ij}x_j,
\end{equation}
where $C_{ij}$ are the elements of an orthogonal matrix $\mathbf{C}$ that diagonalises $\mathbf{K}$,
\begin{equation}
\label{orthogonal}
    \sum_{i,j=1}^d C_{il}\kappa_{ij}C_{jm} = k_l \delta_{lm}, \quad \sum_{i=1}^d C_{il}C_{im} = \delta_{lm}.
\end{equation}
The harmonic potential may be thus expressed in terms of the normal modes,
\begin{equation}
    U(\mathbf{x}) = \frac{1}{2}\sum_{i,j=1}^d\kappa_{ij}\left(\sum_{l=1}^d C_{il} x_l\right)\left(\sum_{m=1}^d C_{jm}x_m\right) = \frac{1}{2}\sum_{l=1}^d k_l x_l^2,
\end{equation}
where we have employed the orthogonality relations from Eq.\eqref{orthogonal}. Thus, we notice that the potential becomes diagonal in the new basis.
On the one hand, we may recover the Fokker-Planck equation~\eqref{Fokker-Planck-equation} by noticing that $P(\mathbf{q},t) = P(\mathbf{x},t)$, since the absolute value of the Jacobian for the coordinate transformation equals unity, and carrying out carefully the change of variables defined by Eq.~\eqref{eq:change-qx}. On the other hand, it is also trivial to proceed similarly for the Langevin equations \eqref{Langevin-equation} and \eqref{Langevin-equation-general}, after identifying
\begin{equation}
    \eta_i(t) = \sum_{j=1}^d C_{ij}\hat{\eta}_j(t),
\end{equation}
as the noise associated to the normal mode $x_i$. The latter corresponds also to a white Gaussian noise, as it satisfies the relations
\begin{eqnarray}
    \langle \eta_i(t) \rangle &=& \sum_{j=1}^d C_{ij} \langle \hat{\eta}_j(t) \rangle = 0,
    \\
    \langle \eta_i(t) \eta_j(t') \rangle &=& \sum_{l=1}^d \sum_{m=1}^d C_{il}C_{jm} \langle \hat{\eta}_l(t) \hat{\eta}_m(t') \rangle = \delta (t-t') \delta_{ij}.
\end{eqnarray}
Once again, the relations from Eq.~\eqref{orthogonal} have been employed.

\section{Pontryagin's maximum principle} 
\label{ap:pontryagin}
In this section, we show how to apply Pontryagin's principle to study the time-optimisation problem of a Brownian particle in a general ($d$-dimensional) harmonic potential. In addition, we will also show the detailed derivation for the two dimensional case, mainly due to its simplicity.

The control problem may be cast in the following way: Let us introduce the control system
\begin{equation}
    \dot{z}_i = f_i(\mathbf{z},\mathbf{k};T) \equiv -2k_iz_i + 2T, \quad i=1,...,d,
\end{equation}
with $\mathbf{z}\equiv (z_1,z_2,...,z_d)$ being the dynamic variables, $\mathbf{k} \equiv (k_1,k_2,...,k_d)$ the stiffness in each dimension and $T = T(t)$ the control parameter of the system. We want to minimise the functional 
\begin{equation}
    J[T] \equiv \int_0^{t_f} dt \ \underbrace{f_0(\mathbf{z},\mathbf{k};T)}_{= 1} = \int_0^{t_f}dt = t_f,
\end{equation}
which corresponds to the final time of the process, given the constraints $T_{\min} \leq T(t) \leq T_{\max}$---i.e. the control $T(t)$ belongs to the control set defined by the interval $[T_{\min},T_{\max}]$, and the boundary conditions
\begin{equation}
    T(0) \equiv T_0 = 1, \quad T(t_f) \equiv T_f, \quad z_i(0) = \frac{T_0}{k_i} = \frac{1}{k_i}, \quad z_i(t_f) = \frac{T_f}{k_i},
\end{equation}
for all $i = 1,...,d$. Now, we define a new variable $z_0$ with $z_0(0) = 0$ such that
\begin{equation}
    \dot{z}_0 = f_0(\mathbf{z},\mathbf{k};T).
\end{equation}
We note that $z_0(t_f)$ corresponds just to $t_f$, which is the magnitude we intend to optimise. Next, we introduce the conjugate variables $(\psi_0, \boldsymbol{\psi}) \equiv (\psi_0,\psi_1,\psi_2,...,\psi_d)$, and the so-called Pontryagin's Hamiltonian 
\begin{equation}
\label{hamiltonian}
    \Pi (\mathbf{z},\psi_0,\boldsymbol{\psi},\mathbf{k};T) \equiv \psi_0 f_0 + \boldsymbol{\psi}\cdot \mathbf{f},
\end{equation}
with $\mathbf{f} \equiv (f_1,f_2,...,f_d)$. In conjunction with Eq.~\eqref{hamiltonian}, the variables $(\mathbf{z},\psi_0,\boldsymbol{\psi})$ satisfy Hamilton's canonical equations
\begin{eqnarray}
    \dot{z}_0 &=& \frac{\partial \Pi}{\partial \psi_0} = f_0, \quad \dot{z}_i = \frac{\partial \Pi}{\partial \psi_i} = f_i, 
    \\
    \dot{\psi}_0 &=& - \frac{\partial \Pi}{\partial z_0} = 0 \ \rightarrow \ \psi_0 = \textnormal{constant},
    \\
    \dot{\psi}_i &=& - \frac{\partial \Pi}{\partial z_i} = 2k_i \psi_i \ \rightarrow \ \psi_i(t) = \psi_{i,0} \ e^{2k_it}.
\end{eqnarray}
Now, Pontryagin's maximum principle states that the Hamiltonian must attain a maximum at the optimal control, implying that
\begin{equation}
    \frac{\partial \Pi}{\partial T} = 2 \sum_{i=1}^d \psi_i = 0.
\end{equation}
However, as the latter expression does not depend on the control parameter $T$, we conclude that the Hamiltonian attains its maximum at the boundaries of the control set. This entails that the solution of the control problem corresponds to a \textit{bang-bang} protocol, for which there are time windows where $T(t) = T_{\max}$ or $T(t) = T_{\min}$. The choice between the maximum and minimum values of the temperature depends on the sign of the derivative of the Hamiltonian. Let us note that, for $d=1$, such derivative reduces to $2\psi_1$, and as $\psi_1$ corresponds to a purely exponential function, it  keeps its sign throughout the entire time window. Thus, the optimal protocol comprises in this case only one time window with either $T(t) = T_{\max}$ or $T(t) = T_{\min}$, depending on whether we intend to heat or cool the system. Now, for the two dimensional case, we have that
\begin{equation}
    \frac{\partial \Pi}{\partial T} = 2\left( \psi_{1,0}\ e^{2k_1t} + \psi_{2,0}\ e^{2k_2 t} \right).
\end{equation}
Thus, assuming that $\textnormal{sign}(\psi_{1,0}) \neq \textnormal{sign}(\psi_{2,0})$, there exists a time $\tau \in [0,t_f]$ for which the derivative of the Hamiltonian changes its sign,
\begin{equation}
    \frac{\partial \Pi}{\partial T} = 0 \quad \iff \quad \tau = \frac{1}{2(k_2-k_1)} \ln \left( -\frac{\psi_{1,0}}{\psi_{2,0}} \right).
\end{equation}

We need to specify the initial values $\psi_{1,0}$ and $\psi_{2,0}$ of the conjugate variables. These  depend on whether the optimal control corresponds to the heating or the cooling cases. Let us focus on the heating one. According to Pontryagin's principle, the Hamiltonian must be zero at the optimal control, for all times. We assume that $T(t=0^+) = T_{\max}$ and $T(t=t_f^-) = T_{\min}$. Thus, we may impose the following equations:
\begin{eqnarray}
    \Pi(0^+) = 0 \ \rightarrow \ 2(T_{\max}-1)(\psi_{1,0} + \psi_{2,0}) = -\psi_0,
    \\
    \Pi(t_f^-) = 0 \ \rightarrow \ 2(T_{\min}-T_f)(\psi_{1,0}\ e^{2k_1t_f} + \psi_{2,0}\ e^{2k_2 t_f}) = -\psi_0,
\end{eqnarray}
from which we obtain
\begin{eqnarray}
    \psi_{1,0} &= \frac{1}{2}\frac{\psi_0}{e^{2k_2 t_f}-e^{2k_1 t_f}}\left[ \frac{e^{2k_2 t_f}}{1-T_{\max}} - \frac{1}{T_f-T_{\min}} \right],
    \\
    \psi_{2,0} &= \frac{1}{2}\frac{\psi_0}{e^{2k_2 t_f}-e^{2k_1 t_f}}\left[  \frac{1}{T_f-T_{\min}} - \frac{e^{2k_1 t_f}}{1-T_{\max}} \right].
\end{eqnarray}
Thus, the switching time $\tau$ is given by
\begin{equation}
    \tau = \frac{1}{2(k_2-k_1)} \ln \left[\frac{1-T_{\max} -(T_f-T_{\min})e^{2k_1t_f}}{1-T_{\max}-(T_f-T_{\min})e^{2k_2t_f}} \right], 
\end{equation}
which is positive for $T_f>1$. We must note that a negative value of $\tau$ would have been obtained had we chosen the opposite order of the bangs, i.e.  $T(t=0^+) = T_{\min}$ and $T(t_f^-) = T_{\max}$. This means that the assumed order of the bangs is indeed the right one for $T_f>1$. In order to fully determine the unknown parameters $(\tau,t_f)$, we would need to resort to one of the evolution equations given by Eq.~\eqref{two-dim-heating} (either with $k = k_1$ or $k = k_2$), where we identify $\tau = \tau_1$ and $t_f = \tau_1 + \tau_2$.

\section{Unphysical scenario: Negative temperatures}
\label{ap:Tmin-inf}
Let us consider the two-dimensional harmonic oscillator, for general values of the boundary temperatures $T_{\max}$ and $T_{\min}$ and for the heating control, $T_f>1$, in the $k_2 \rightarrow k_1=1$ limit. Direct resolution of the evolution equations \eqref{eq:comp-deg1} over the whole time window leads to the equations
\begin{eqnarray}
    T_f &=& (1-T_{\max})e^{-2(\tau_1+\tau_2)} + \left(T_{\max}-T_{\min}\right)e^{-2\tau_2} +T_{\min},
    \\
    0 &=& (\tau_1+\tau_2)\left(1-T_{\max}\right)e^{-2(\tau_1+\tau_2)} + \tau_2\left(T_{\max}-T_{\min}\right)e^{-2\tau_2}.
\end{eqnarray}
Now, for this unphysical scenario, we assume that $T_c \equiv T_{\max} = - T_{\min}$. This assumption implies that we can heat up the system as much as we can cool it down. We would recover a sort of ``temperature symmetry" that it is not actually present in the physical case. Within this assumption, the evolution equations reduce to
\begin{eqnarray}
\label{reduced-evol-eqs-Tc-1}
    T_f &=& (1-T_{c})e^{-2(\tau_1+\tau_2)} + 2T_ce^{-2\tau_2} - T_{c},
    \\
    \label{reduced-evol-eqs-Tc-2}
    0 &=& (\tau_1+\tau_2)(1-T_c)e^{-2(\tau_1+\tau_2)} + 2\tau_2T_ce^{-2\tau_2},
\end{eqnarray}
\begin{figure}
\flushright\includegraphics[width=0.875\textwidth]{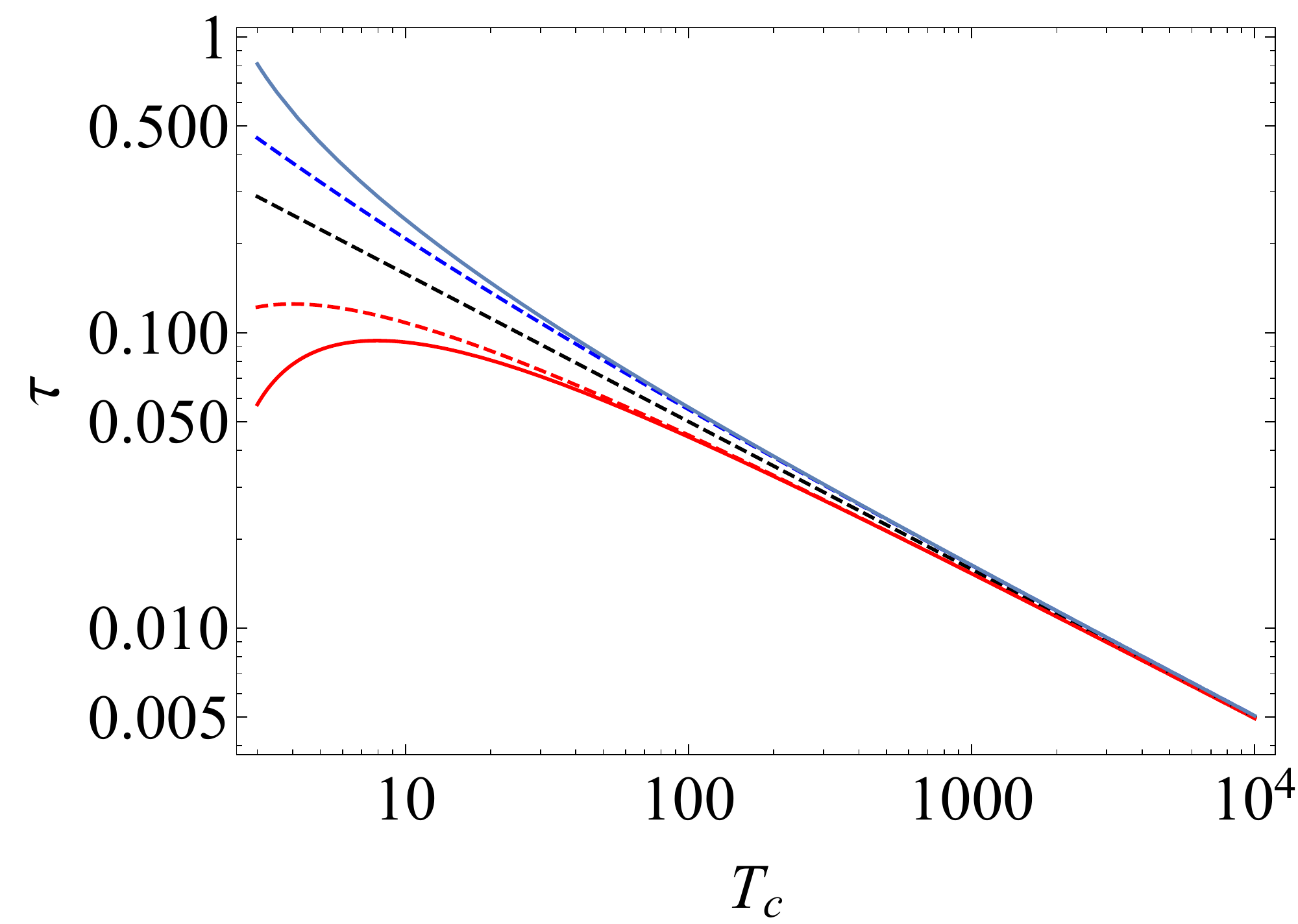}
\caption{Boundary temperature dependence of $\tau_1$ (blue) and $\tau_2$ (red) for the heating protocol in a two dimensional unphysical harmonic oscillator, for which $T_{\min}=-T_{\max}=T_c$. We consider the almost fully degenerate case $k_2 \rightarrow k_1 = 1$, for $T_f =2$. Full lines correspond to the numerical solution of Eqs.~\eqref{reduced-evol-eqs-Tc-1} and \eqref{reduced-evol-eqs-Tc-2}, while the black dashed line corresponds to the $O(T_c^{-1/2})$ contribution for both time intervals, and the blue and red dashed lines, respectively, to $\tau_1$ and $\tau_2$ up to $O(T_c^{-1})$. All dashed curves are analytical and stem from Eq.~\eqref{times-unphysical}.} \label{fig:times-Tc}
\end{figure}

\begin{figure}
\flushright
\includegraphics[width=0.875\textwidth]{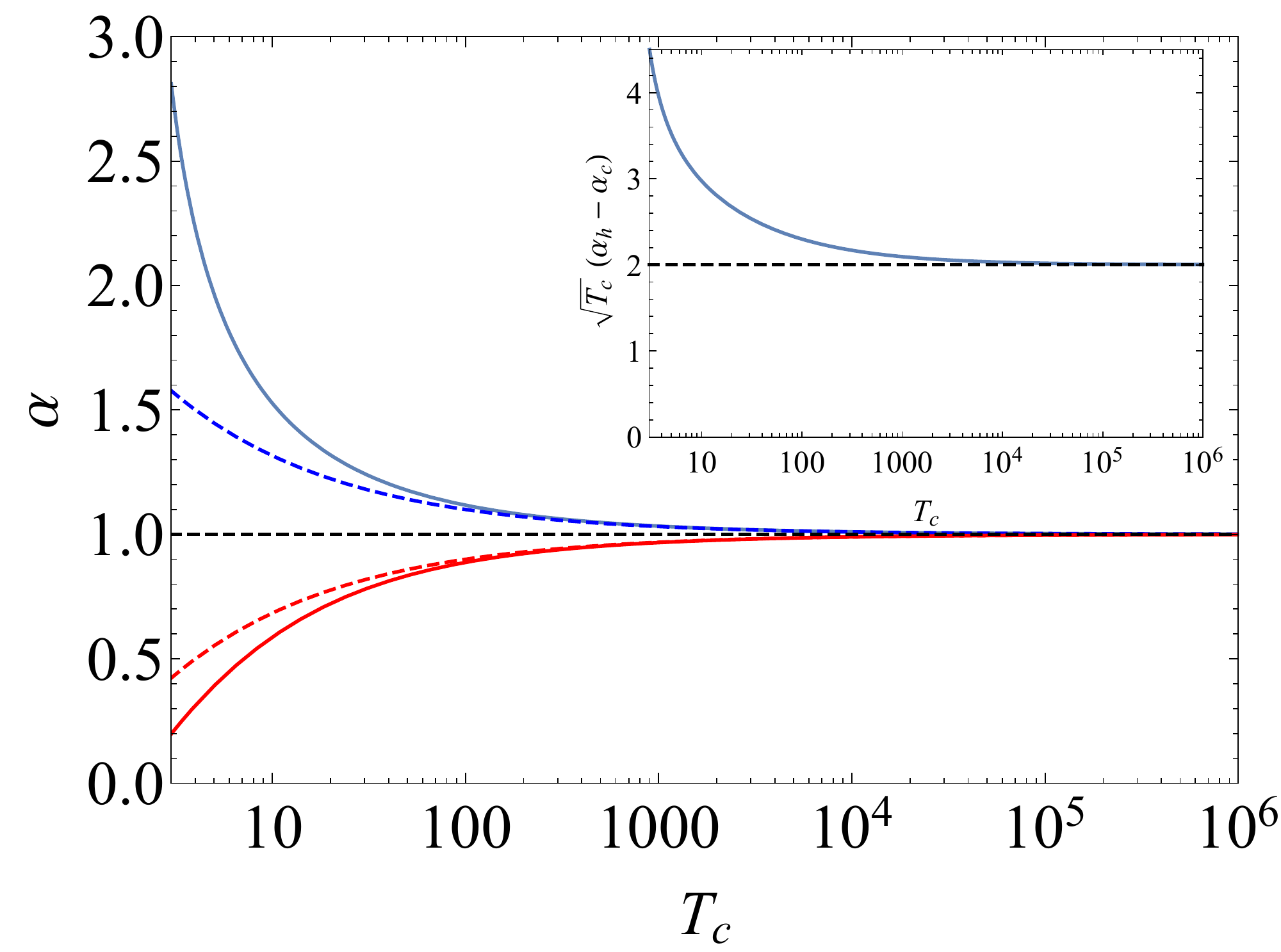}
\caption{Boundary temperature dependence of the constants $\alpha_h$ (blue) and $\alpha_c$ (red) for the heating protocol in a two dimensional unphysical harmonic oscillator.  As in Fig.~\ref{fig:times-Tc}, $k_2 \rightarrow k_1 = 1$ and $T_f =2$. Full lines correspond to the numerical solution of the evolution equations Eqs.~\eqref{reduced-evol-eqs-Tc-1} and \eqref{reduced-evol-eqs-Tc-2}, plus the definitions from Eq.~\eqref{alpha-constants}, while the black dashed line corresponds to the $O(1)$ contribution for both constants, and the blue and red dashed lines, respectively, to $\alpha_h$ and $\alpha_c$ up to $O(T_c^{-1/2})$. All dashed curves are  analytical predictions stemming from Eqs.~\eqref{alpha-0} and \eqref{alpha-1}.}\label{fig:alpha-Tc}
\end{figure}

We are interested in the asymptotic behaviour of both $\tau_1$ and $\tau_2$ when $T_c \rightarrow +\infty$. Such infinite power to both heat and cool the system entails vanishing times for both the heating and the cooling windows. However, as Figure \ref{fig:times-Tc} shows, the asymptotic behaviour of both $\tau_1$ and $\tau_2$ scales as $T_c^{-1/2}$ in this case. Thus, it is appealing to introduce the constants
\begin{eqnarray}
\label{alpha-constants}
    \alpha_{h} = 2\sqrt{T_c}\tau_1, \quad \alpha_{c} = 2\sqrt{T_c}\tau_2.
\end{eqnarray}
For $T_c \rightarrow +\infty$, both $\tau_1$ and $\tau_2$ may be expanded in powers of $T_c^{-1/2}$ as
\begin{eqnarray}
    \tau_1 = \frac{\tau_1^{(0)}}{\sqrt{T_c}} + \frac{\tau_1^{(1)}}{T_c} + \frac{\tau_1^{(2)}}{T_c^{3/2}}  \cdots  \ \Rightarrow \ \alpha_h =  \underbrace{2 \tau_1^{(0)}}_{\equiv \alpha_h^{(0)}} + \underbrace{\frac{2\tau_1^{(1)}}{\sqrt{T_c}}}_{\equiv \alpha_h^{(1)}/\sqrt{T_c}} + \underbrace{\frac{2\tau_1^{(2)}}{T_c}}_{\equiv \alpha_h^{(2)}/T_c} \cdots,
    \\
    \tau_2 = \frac{\tau_2^{(0)}}{\sqrt{T_c}} + \frac{\tau_2^{(1)}}{T_c}  + \frac{\tau_2^{(2)}}{T_c^{3/2}}  \cdots \ \Rightarrow \ \alpha_c =  \underbrace{2 \tau_2^{(0)}}_{\equiv \alpha_c^{(0)}} +  \underbrace{\frac{2\tau_2^{(1)}}{\sqrt{T_c}}}_{\equiv \alpha_c^{(1)}/\sqrt{T_c}} + \underbrace{\frac{2\tau_2^{(2)}}{T_c}}_{\equiv \alpha_c^{(2)}/T_c} \cdots.
\end{eqnarray}
With these definitions, $T_c \rightarrow +\infty$ in the reduced evolution equations leads to
\begin{eqnarray}
    T_f &=& 1 + \sqrt{T_c}(\alpha_h^{(0)} - \alpha_c^{(0)}) +(\alpha_h^{(1)} - \alpha_c^{(1)})
    \nonumber
    \\
    &&-\frac{1}{2}(\alpha_h^{(0),2} + 2\alpha_h^{(0)} \alpha_c^{(0)} - \alpha_c^{(0),2}) + O(T_c^{-1/2}), 
    \\
    0 &=& \sqrt{T_c}(\alpha_h^{(0)} - \alpha_c^{(0)}) +(\alpha_h^{(1)} - \alpha_c^{(1)})
    \nonumber
    \\
    &&-(\alpha_h^{(0),2} + 2\alpha_h^{(0)} \alpha_c^{(0)} - \alpha_c^{(0),2}) + O(T_c^{-1/2}), 
\end{eqnarray}
On the one hand, the above system of equations is only consistent if $\alpha_h^{(0)}$ and $\alpha_c^{(0)}$ are equal, such that the $O(\sqrt{T_c})$ terms vanish. On the other hand, we are left with an undetermined system of two equations for the variables $(\alpha_h^{(0)},\alpha_h^{(1)},\alpha_c^{(1)})$,
\begin{eqnarray}
    \label{alpha-0}
    T_f &=& 1 + \alpha_h^{(0),2},
    \\
    \label{alpha-0-2}
    0 &=& \alpha_h^{(1)} - \alpha_c^{(1)} - 2\alpha_h^{(0),2}.
\end{eqnarray}
In order to close the system, we need to take into account the subdominant contributions in $T_c$. Those of $O(T_c^{-1/2})$ read
\begin{eqnarray}
    0 &=& \alpha_h^{(0),3} - 2\alpha_h^{(0)}(1+\alpha_h^{(1)}) + (\alpha_h^{(2)}-\alpha_c^{(2)}),
    \\
    0 &=& 3\alpha_h^{(0),3} - 2\alpha_h^{(0)}(1+2\alpha_h^{(1)}) + (\alpha_h^{(2)}-\alpha_c^{(2)}),
\end{eqnarray}
from which we obtain 
\begin{equation}
\label{alpha-1}
    \alpha_h^{(1)} = \alpha_h^{(0),2}, \ \alpha_c^{(1)} = -\alpha_h^{(0),2}.
\end{equation}
In general, the $O(T_c^{(1-n)/2})$ allows us to solve for the variables $\alpha_h^{(n)}-\alpha_c^{(n)}$ and $\alpha_h^{(n-1)}$. Finally, up to $O(T_c^{-1})$, the time intervals are given by
\begin{equation}
\label{times-unphysical}
    \tau_1 = \frac{1}{2}\sqrt{\frac{T_f-1}{T_c}} + \frac{T_f-1}{T_c}, \quad \tau_2 = \frac{1}{2}\sqrt{\frac{T_f-1}{T_c}} - \frac{T_f-1}{T_c},
\end{equation}
which clearly vanish in the $T_c \rightarrow +\infty$ limit. In Figs.~\ref{fig:times-Tc} and \ref{fig:alpha-Tc}, the excellent agreement between our asymptotic analysis (dashed lines) and the numerical solution (solid lines) is evident.




\end{appendices}


\bibliography{Mi-biblioteca-21-jul-2022}


\end{document}